\documentclass[%
 aip,
 amsmath,amssymb,
 reprint,%
]{revtex4-1}

\usepackage{graphicx}
\usepackage{xcolor}
\usepackage{ulem}
\usepackage[final]{changes}
\usepackage{dcolumn}
\usepackage{bm}

\usepackage[utf8]{inputenc}
\usepackage[T1]{fontenc}
\usepackage{mathptmx}
\usepackage{etoolbox}

\makeatletter
\def\@email#1#2{%
 \endgroup
 \patchcmd{\titleblock@produce}
  {\frontmatter@RRAPformat}
  {\frontmatter@RRAPformat{\produce@RRAP{*#1\href{mailto:#2}{#2}}}\frontmatter@RRAPformat}
  {}{}
}%
\makeatother
\begin{document}

\preprint{AIP/123-QED}

\title{The elasticity of semiflexible polymers with reversible spontaneous curvature \added{in two dimensions}}
\author{Donghyeon Kim, Panayotis Benetatos}
 \email{pben@knu.ac.kr}
\affiliation{Department of Physics, Kyungpook National University, 80 Daehakro, Bukgu, Daegu, 41566, Republic of Korea
}%

\date{\today}

\begin{abstract}
Many semiflexible filaments exhibit structural asymmetries that lead to a curved ground state (curvature in the absence of thermal fluctuations, external forces, or torques). This property is known as spontaneous curvature. Many semiflexible polymers can reversibly develop spontaneous curvature through internal conformational transitions or through transient interactions with their molecular environment. In this article, we show that such reversible spontaneous curvature, even in the absence of transitions in the value of the bending stiffness, provides a minimal mechanism for ensemble-dependent elasticity. We formulate an analytically tractable two-state modified wormlike-chain model in two dimensions, in which an uncurved state (without \added{spontaneous} \deleted{spantaneous} curvature) competes with a curved state (possessing spontaneous curvature). The transition is controlled by an activation energy. Within the weak-bending approximation, we obtain the Gibbs and Helmholtz partition functions for a stretched filament with reversible constant (uniform along the backbone) or sinusoidal spontaneous curvature and for a grafted bistable filament driven \added{either} by an end torque \added{or a bending force}. Stretching \added{or bending} induces a crossover from spontaneous-curvature-dominated to entropic elasticity. We show how the elastic response differs in the two ensembles (Gibbs vs. Helmholtz). In addition, we consider the case where the curved state is also characterized by a higher value of the bending stiffness. In that case, a reentrant transition is possible as we stretch the bistable filament. 

\end{abstract}

\maketitle

\section{\label{sec:Intro}Introduction}

The use of statistical mechanics to explain the elasticity of polymers at the single molecule level goes back to Flory~\cite{Florybookchains}. However, contrary to the usual treatment of macroscopic systems at the thermodynamic limit, where the same thermodynamic behavior emerges irrespective of the statistical ensemble~\cite{Callen}, polymers quite often exhibit statistical ensemble inequivalence. This refers to different stress-strain relations, depending on which parameter is fixed (controlled) and which is fluctuating (averaged). The reason is the unattainability of the thermodynamic limit under certain conditions and the related persistence of strong fluctuations. For single-state flexible macromolecules described as Gaussian chains, there is ensemble equivalence, provided that the elastic response is expressed in terms of conjugate variables~\cite{WinklerSoft,neuman2008single,NeumannBPJ2003,NeumannPRA1985,NeumannPRA1986,Holm,Skvortsovreview2012}. For semiflexible polymers whose conformations are dominated by the persistence length, it is known that there is ensemble inequivalence~\cite{SinhaSamuel,triple_minima}. This is particularly important for the proper interpretation of single-molecule elasticity experiments.

There are many different mechanisms that can give rise to ensemble inequivalence in the elastic response of single polymers. Semiflexibility is one of them. Confinement is another mechanism that can yield ensemble inequivalence even in flexible polymers~\cite{PB_Dutta_SM1,PB_Dutta_SM2,Skvortsov_JCP_2007,Skvortsov_JSMTE_2004,Skvortsov_Macromol_Symp_2006}. A broad category that also exhibits ensemble inequivalence is that of two-state or multi-state polymers. In these polymers, the observed macrostate is an average over two or more distinct coarse-grained descriptions. The effective free energy conformational functional, which we call "Hamiltonian" in polymer physics, involves a binary or higher-order discrete degree of freedom corresponding to the various metastable states. Examples include polymers with extensible units or bonds~\cite{Radiom_Borkovec_PRE,Giordano_Continuum_Mechanics,Giordano_Inventions,Giordano_JCP_bistableFJC,Giordano_JCP_2012,Giordano_Physica_A_2014,Giordano_PRE_2018,Giordano_PRE_two-state,Muscle_PRL,Muscle_PRE,Giordano_SM_2017}, freely jointed chains with reversible hinges~\cite{Reineker,rFJC_SM,rFJC_JCP}, wormlike chains with two-valued reversible bending stiffness~\cite{benetatos2022ensemble,razbin2023elasticity}, flexible polymers undergoing reversible looping or zipping~\cite{Noh_PB_loop_PRE}, and rotaxane-based daisy chains~\cite{Giordano_daisy_2025}. When these systems have finite size, they all exhibit ensemble inequivalence. 

In this article, we investigate a different mechanism that can give rise to nontrivial elasticity and ensemble inequivalence in semiflexible polymers. We consider the case of reversible spontaneous curvature. Spontaneous (or intrinsic) curvature refers to the curvature of the polymer backbone in the absence of thermal fluctuations (ground state) or any external force or torque. This reversible switching of spontaneous curvature can happen through internal conformational changes or through the attachment/detachment of other macromolecules from the environment of the original polymer. Bacterial flagella provide an example of all-or-none structural transitions between states of different spontaneous curvature~\cite{Srigiriraju_PRE_2006,Vogel_PRL_2013}. FtsZ filaments are the bacterial homologue of microtubules. They also undergo a structural transition from a mostly straight state to a curved state~\cite{Li_Ftsz_Science_2013,Erickson_FtsZ_PNAS_2009}. An RNA-DNA hybrid duplex has higher spontaneous curvature than ssDNA~\cite{Wu_RDH_2025}. These duplexes are formed reversibly~\cite{Sugimoto_RDH_thermodynamics_1995,Watkins_RDH_oligo_2005}. Reversible spontaneous curvature can appear locally on DNA with the attachment/detachment of specific proteins, such as IHF~\cite{Swinger_IHF_2004}, TBP~\cite{Nikolov_TBP_1994}, HMG-box~\cite{DRAGAN_HMG_Box_2004}, and CAP/CRP~\cite{Schultz_CAP_1991}.

The emphasis of the present study is on the qualitative features of the elastic response of polymers with reversible spontaneous curvature. To this end, we build on the two-dimensional model of Ref.~\onlinecite{benetatos2010stretching}. In the weakly bending approximation, which is suitable for strong stretching, one obtains a closed analytic force-extension relation, where the thermal part and the curvature-dominated part decouple.

\added{While the present analysis addresses equilibrium elasticity rather than kinetics, we should point out that our models are based on the assumption of all-or-none transitions between distinct conformational states. That implies that the kinetics reduces to crossing between two free energy minima. An expansion of our work in the time domain would involve contact with significant previous works on force-dependent free energy landscapes~\cite{DudkoPRL2006,ZacconePRE2017}.}

This article is organized as follows. In Sec.~\ref{sec:SP_const}, we calculate analytically the force-extension response and the probability of occurrence of the curved state as a function of the relevant control parameter, in both the Gibbs and the Helmholtz ensemble, for a two-state semiflexible polymer with reversible constant spontaneous curvature. We also present a convenient approximation that simplifies the results. In Sec.~\ref{sec:SP_sin}, we go through the same analysis, for a two-state semiflexible polymer with reversible sinusoidal spontaneous curvature along its contour. The exact results here require a numerical step, but there is also an analytic approximation. In Sec.~\ref{sec:TA} \added{and ~\ref{sec:FY}}, we analyze, in both ensembles, the torque-angle \added{and transverse force-displacement} elastic response of a two-state grafted rodlike polymer with reversible spontaneous curvature. We summarize in Sec.~\ref{sec:Conclusions}. Some details of the calculations are shown in the Appendices. 

\added{A note on the notation: In the main part of this article, we consider two fixed states as well as the mixed state. The state without spontaneous curvature is state 0 and the corresponding persistence length is denoted by $L_P^{(0)}$. The state with constant spontaneous curvature is state 1 and the corresponding persistence length is denoted by $L_P^{(1)}$. The state with sinusoidal spontaneous curvature is state 2 and the corresponding persistence length is denoted by $L_P^{(2)}$.}

\section{\label{sec:SP_const} Polymer under tension with reversible constant spontaneous curvature}
We first consider a stretched semiflexible polymer whose curved state is characterized by a constant spontaneous curvature, that is, the same spontaneous curvature along its backbone. As a starting point, we briefly recall the modified wormlike-chain model introduced in Ref.~\onlinecite{benetatos2010stretching} for a two-dimensional weakly bending filament with a prescribed spontaneous-curvature profile. \added{Irrespective of the value of the spontaneous curvature, the weak-bending approximation can be realized in a strong-stretching regime.} In that model, the filament is represented by a locally inextensible plane curve \(\mathbf r(s)\), where \(s\in[0,L]\) is the arc length and
\begin{align}
\left|\frac{\partial \mathbf r(s)}{\partial s} \right|=1 .
\end{align}
The tangent vector is \(\mathbf t(s)=\partial_s\mathbf r(s)\), and the effective free-energy functional that penalizes bending is written as
\begin{align}
H_0[\mathbf r]
= \frac{\kappa}{2}\int_0^L ds\,
\left[
\left|\frac{\partial \mathbf t(s)}{\partial s}\right|-c(s)
\right]^2 ,
\label{eq:modified_wlc}
\end{align}
where \(\kappa\) is the bending stiffness (units of energy $\times$ length) and \(c(s)\) denotes the spontaneous curvature along the contour. Thus, the elastic cost is determined by the deviation of the local curvature from its preferred value.

For a filament stretched by a force \(f\) along the \(x\)-direction, Ref.~\onlinecite{benetatos2010stretching} considers the weak-bending limit, in which the angle that the tangent vector makes with the $x$-axis is small. To quadratic order in that angle, the corresponding Hamiltonian can be written as
\begin{align}
\label{eq:weak_bending_hamiltonian}
H_f[y]= \frac{\kappa}{2}\int_0^L ds\,
\left[\frac{\partial^2 y(s)}{\partial s^2}-\tilde c(s)\right]^2 \nonumber\\
+ \frac{f}{2}\int_0^L ds\,\left[\frac{\partial y(s)}{\partial s}\right]^2-fL . 
\end{align}
Here \(\tilde c(s)\) is the signed spontaneous curvature in two dimensions; its sign distinguishes the two bending directions. 

Minimizing Eq.~\eqref{eq:weak_bending_hamiltonian} requires four boundary conditions. Following Ref.~\onlinecite{benetatos2010stretching}, we use hinged--hinged conditions,
\begin{align}
y(0)=0, \qquad \partial_s^2 y(0)=\tilde c(0),\nonumber\\
y(L)=0, \qquad \partial_s^2 y(L)=\tilde c(L). 
\label{eq:hinged_boundary_conditions}
\end{align}
These conditions place both end points on the pulling axis and correspond to torque-free ends. In the strong-stretching regime, however, the leading force--extension response is insensitive to the detailed boundary conditions. 

In this section, the curved state is taken to have a constant signed spontaneous curvature, \(\tilde c(s)=c_0\), while the uncurved state corresponds to \(\tilde c(s)=0\). A sinusoidal profile, \(\tilde c(s)=c_0\sin(qs)\), is considered in the Sec.~\ref{sec:SP_sin}. The activation energy of the curved state is introduced below in the two-state partition function.

\begin{figure}[h]
\centering
\includegraphics[width=0.4\textwidth]{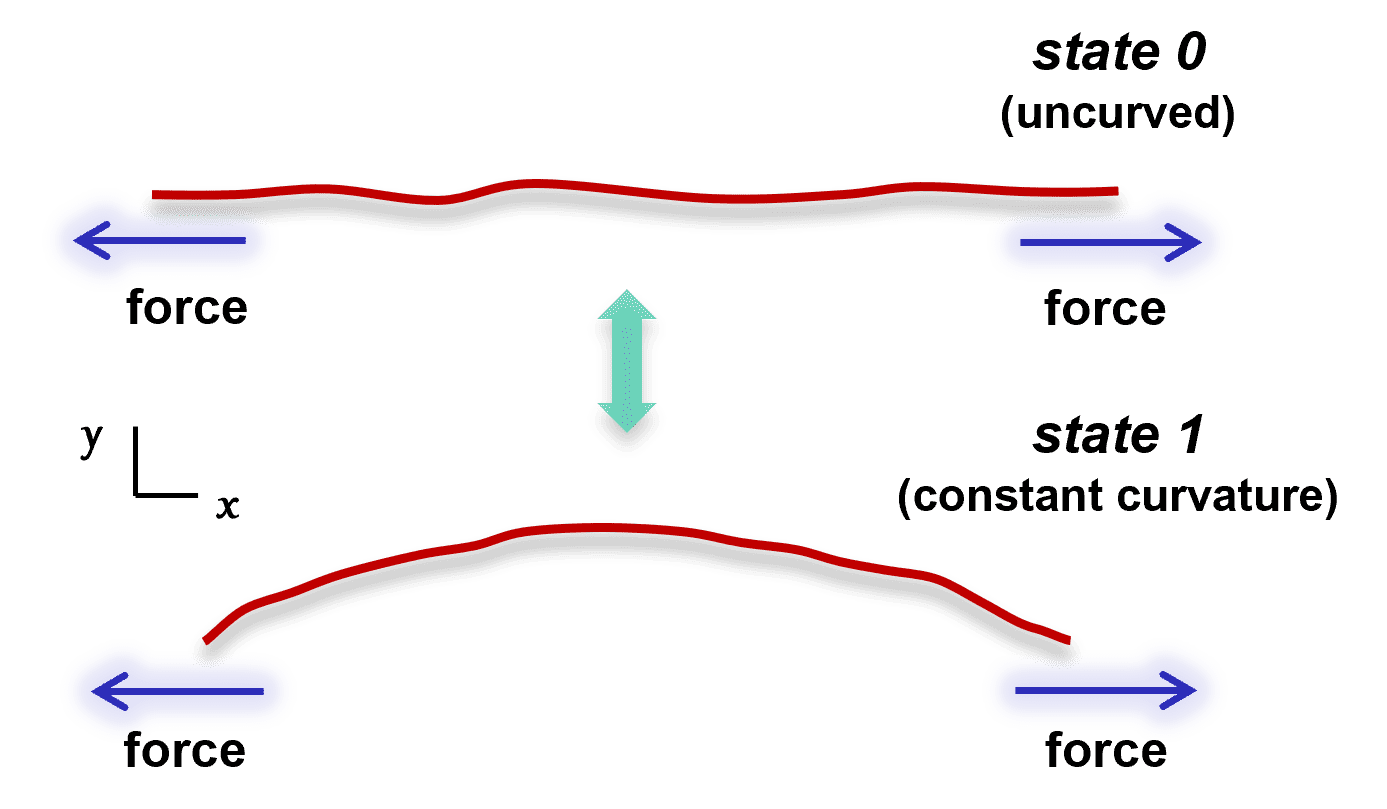}
\caption{Schematic representation of the two states (uncurved vs. curved). The polymer fluctuates between these two states.}
\label{fig:SP_intro}
\end{figure}

For a filament with constant spontaneous curvature, we use the strong-stretching force--extension relation of Ref.~\onlinecite{benetatos2010stretching} as the single-state response of the curved regime. For \(f\gg f_L\equiv \kappa/L^2\), this relation is
\begin{eqnarray}
\label{sp_const}
    \frac{\langle x \rangle}{L} &= 1 -
\underbrace{\frac{1}{4}\left(\frac{L}{c_0^{-1}}\right)\left(\frac{f_c}{f}\right)^{3/2}}_{\added{\mathrm{curvature\;contribution}}}
-\underbrace{\frac{1}{2}\left(\frac{f_{cr}}{f}\right)^{1/2}}_{\added{\mathrm{thermal\;fluctuation}}} \\
&=1-a f^{-3/2} - b f^{-1/2}. \nonumber
\end{eqnarray}
Here \(c_0\) is the constant spontaneous curvature, \(c_0=1/R\), and the relevant force scales are
\begin{eqnarray}
f_c \equiv \frac{\kappa}{c_0^{-2}},
\qquad
f_{cr}\equiv \frac{\kappa}{L_p^2}.
\end{eqnarray}
For later use, we have introduced
\begin{eqnarray}
a\equiv
\frac{L}{4c_0^{-1}}f_c^{3/2}
=2^{-7/2}Lc_0^4L_p^{3/2},\\
b\equiv
\frac{1}{2}f_{cr}^{1/2}
=2^{-3/2}L_p^{-1/2}. 
\end{eqnarray}
In two dimensions, the persistence length is related to the bending rigidity by
\begin{eqnarray}
L_p=\frac{2\kappa}{k_BT(d-1)}
=\frac{2\kappa}{k_BT}.
\end{eqnarray}
Since \(d=2\) and we use units in which \(k_BT=1\), this reduces to \(L_p=2\kappa\).

The force scale \(f_L=\kappa/L^2\) marks the crossover above which the force-induced orientational correlation length~\cite{Odijk,deGennes2} $\sqrt{\kappa/f}$ is smaller than the total contour length, so that finite-length corrections do not enter
the leading strong-stretching form. The first deviation from full stretching in Eq.~(\ref{sp_const}) is the athermal contribution associated with the spontaneous curvature (straightened by the tensile force). The second deviation is the entropic contribution from thermal transverse fluctuations, identical to the Marko--Siggia strong-stretching term~\cite{marko1995stretching}. In the two-state model below, Eq.~(\ref{sp_const}) is assigned to the curved state, whereas the uncurved-state elasticity is obtained by setting the curvature contribution to zero while retaining the thermal fluctuation term.

We use Eq.~(\ref{sp_const}) to construct a two-state description, as shown in Fig.~\ref{fig:SP_intro}. State \(1\) denotes the curved state and contains both the athermal curvature contribution and the thermal fluctuation contribution. State \(0\) denotes the uncurved state, for which only the thermal fluctuation contribution remains. Thus, the extension of state \(1\) is written as
\begin{eqnarray}
\langle x \rangle_1(f) = L - \frac{La}{f^{3/2}} - \frac{L b_1}{f^{1/2}} .
\label{sp_const_Gc}
\end{eqnarray}
Using the thermodynamic relation \(\langle x\rangle=-\partial G/\partial f\), the corresponding Gibbs free energy is
\begin{eqnarray}
G_1(f) = -fL - \frac{2La}{f^{1/2}} + 2Lb_1f^{1/2}.
\end{eqnarray}

State \(0\) represents a filament without spontaneous curvature ("uncurved"). Its stretching response is governed solely by thermal transverse fluctuations, and is given by
\begin{eqnarray}
\langle x \rangle_0(f) = L - \frac{L b_{0}}{f^{1/2}}.
\label{sp_const_Gth}
\end{eqnarray}
The Gibbs free energy of this branch is
\begin{eqnarray}
G_{0}(f)=-fL + 2Lb_{0}f^{1/2}.
\end{eqnarray}
When the bending stiffness is unchanged between the two states, one has \(b_0=b_1\). We keep the notation \(b_0\) and \(b_1\) to also allow comparison with the case in which the persistence length changes between states. Also, \(a\) and \(b_1\) are evaluated with the same persistence length, since both the curvature and thermal contributions belong to state \(1\).

The curved state is assigned an activation energy \(\epsilon\), which accounts for the energetic cost of forming or stabilizing the spontaneous curvature. The Gibbs partition function is then
\begin{eqnarray}
Z_{01}^G(f)
=\exp\left[-\beta G_0\right]
+\exp\left[-\beta\left(G_1+\epsilon\right)\right],
\end{eqnarray}
where \(\beta=1/k_BT\). The two-state Gibbs free energy is
\begin{eqnarray}
G_{01}(f)=-k_BT\ln Z_{01}^G(f).
\end{eqnarray}

The same partition function gives the mean occupation of the curved state,
\begin{eqnarray}
\langle n_1 \rangle_G(f)
=\frac{1}{\exp\left[\beta\left(G_1-G_0+\epsilon\right)\right]+1}.
\end{eqnarray}
The average extension in the Gibbs ensemble follows from \(\langle x\rangle=-\partial G/\partial f\), or equivalently from the branch-weighted average,
\begin{eqnarray}
\langle x \rangle_{01}(f)
=\left(1-{\langle n_1 \rangle_G}\right)
\langle x \rangle_0
+\langle n_1 \rangle_G
\langle x \rangle_1.
\end{eqnarray}

To formulate the Helmholtz description, we invert the force--extension relation of state \(1\). Defining \(\delta\equiv 1-x/L\) and \(u=f^{-1/2}\), Eq.~\eqref{sp_const_Gc} becomes
\begin{eqnarray}
au^3+b_1u - \delta=0.
\end{eqnarray}
The cubic equation has a unique positive real root for \(\delta>0\), since \(a>0\) and \(b_1>0\). Using Cardano's formula~\cite{abramowitz1948handbook}, this root is
\begin{eqnarray}
u(x)=\left[\frac{\delta}{2a}+\sqrt{\left(\frac{\delta}{2a}\right)^2+\left(\frac{b_1}{3a}\right)^3}\right]^{1/3} \nonumber\\
+\left[\frac{\delta}{2a}-\sqrt{\left(\frac{\delta}{2a}\right)^2+\left(\frac{b_1}{3a}\right)^3}\right]^{1/3}. 
\end{eqnarray}
The corresponding force is 
\begin{eqnarray}
\langle f \rangle_1(x)=u^{-2}(x) .
\label{eq:sp_const_f1}
\end{eqnarray}
The Helmholtz free energies of the two states (branches) are obtained by applying a branchwise Legendre transformation,
\begin{eqnarray}
F_i(x)=G_i(f_i)+f_i x,
\qquad
i\in\{0,1\},
\end{eqnarray}
where \(f_i\) is the force corresponding to the fixed extension \(x\) in branch \(i\). This gives
\begin{eqnarray}
F_1(x)
=-3La\,u(x)+\frac{Lb_1}{u(x)} .
\label{eq:F1_full_curved}
\end{eqnarray}
The derivation of Eq.~\eqref{eq:F1_full_curved} \added{is} \deleted{are} given in Appendix~\ref{appendix:helmholtz_branch}.

For state \(0\), the inversion of Eq.~\eqref{sp_const_Gth} gives
\begin{eqnarray}
\langle f\rangle_0(x)=b_0^2\left(1-\frac{x}{L}\right)^{-2}.
\end{eqnarray}
The corresponding Helmholtz free energy is
\begin{eqnarray}
F_0(x)=Lb_0^2\left(1-\frac{x}{L}\right)^{-1}.
\end{eqnarray}

The Helmholtz partition function of the two-state system is then
\begin{eqnarray}
Z_{01}^H(x)
=\exp\left[-\beta F_0\right]
+\exp\left[-\beta\left(F_1+\epsilon\right)\right].
\end{eqnarray}
The probability of occupying the curved state in the Helmholtz ensemble is
\begin{eqnarray}
\langle n_1 \rangle_H(x)
=\frac{1}{\exp\left[\beta\left(F_1-F_0+\epsilon\right)\right]+1}.
\end{eqnarray}

The Helmholtz free energy of the two-state system is
\begin{eqnarray}
F_{01}(x)=-k_BT\ln Z_{01}^H(x).
\end{eqnarray}
Using the thermodynamic relation \(\langle f\rangle=\partial F/\partial x\), the mean force is obtained as the occupation-weighted average of the two branch forces:
\begin{eqnarray}
\langle f \rangle_{01}(x)
=\left(1-\langle n_1 \rangle_H\right)\langle f \rangle_0
+\langle n_1 \rangle_H\langle f \rangle_1 .
\end{eqnarray}

\begin{figure}[h]
\centering{\includegraphics[width=0.5\textwidth]{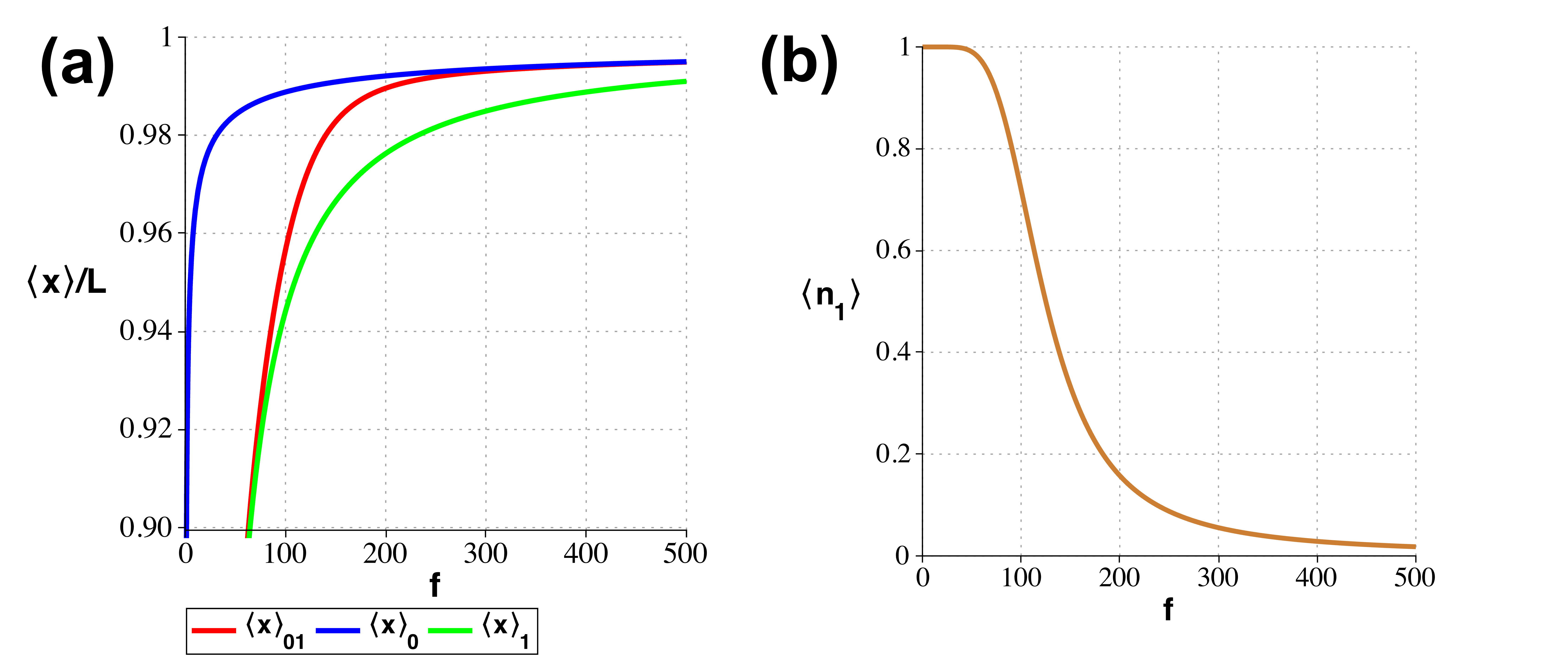}}
\caption{Force--extension response of a two-state polymer with switching  spontaneous curvature but fixed bending stiffness, in the Gibbs (fixed force) ensemble. (a) Force–extension curves. The red curve shows the two-state response $\langle x \rangle_{01}$, the blue and green correspond to the uncurved state $\langle x \rangle_0$ and the curved state $\langle x \rangle_1$, respectively. (b) Probability of the curved state, $\langle n_1 \rangle_G$, as a function of the applied force $f$. The calculation uses the following parameters: $L=1$, $\beta=1$, $\epsilon=8$, $c_0=2$, \(L_p^{(0)}=L_p^{(1)}=10\).}
\label{fig:SP_c_G}
\end{figure}
\begin{figure}[h]
\centering{\includegraphics[width=0.5\textwidth]{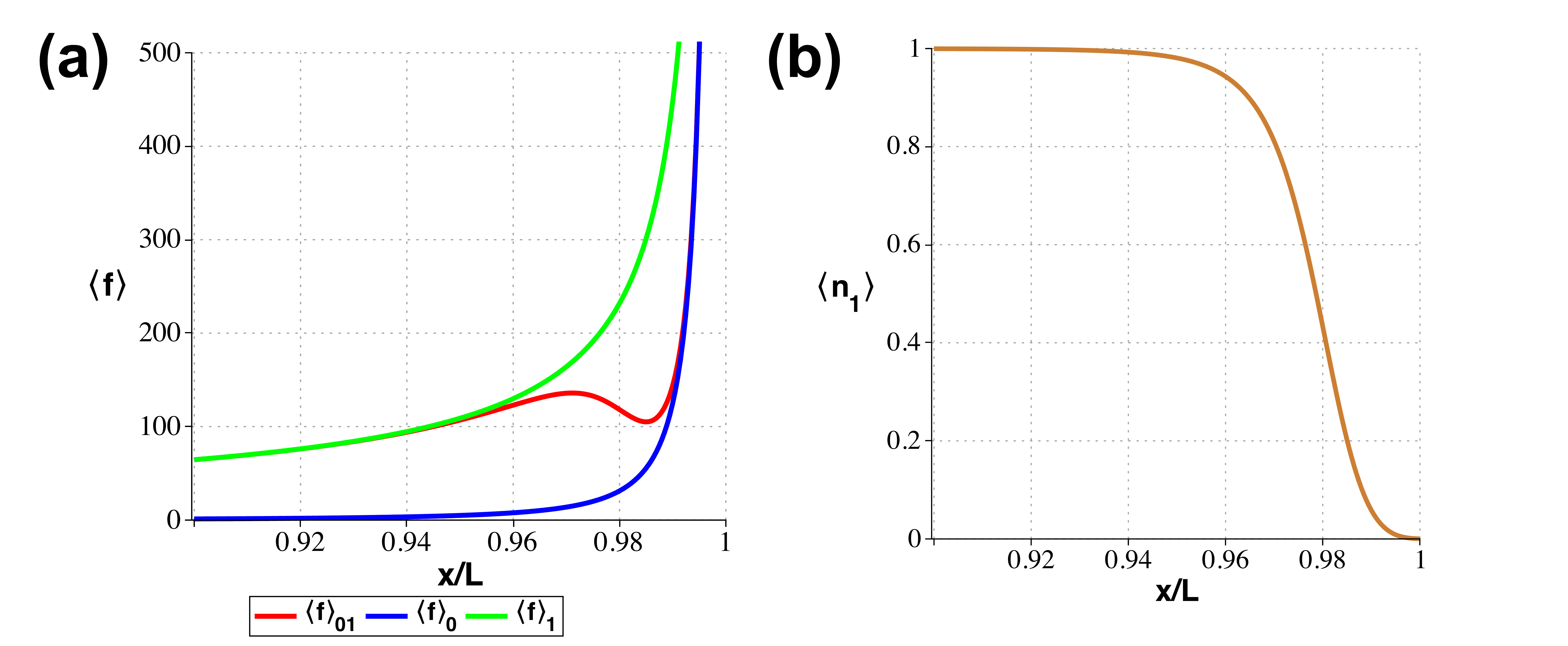}}
\caption{Force--extension response of a two-state polymer with switching  spontaneous curvature but fixed bending stiffness, in the Helmholtz (fixed extension) ensemble. (a) Force–extension curves. The red curve shows the two-state response $\langle f \rangle_{01}$, the blue and green correspond to the uncurved state $\langle f \rangle_0$ and the curved state $\langle f \rangle_1$, respectively. (b) Probability of the curved state, $\langle n_1 \rangle_H$, plotted against the normalized extension $x/L$. The parameters are the same as in Figure~\ref{fig:SP_c_G}: $L=1$, $\beta=1$, $\epsilon=8$, $c_0=2$, \(L_p^{(0)}=L_p^{(1)}=10\).}
\label{fig:SP_c_H}
\end{figure}
\begin{figure}[h]
\centering{\includegraphics[width=0.5\textwidth]{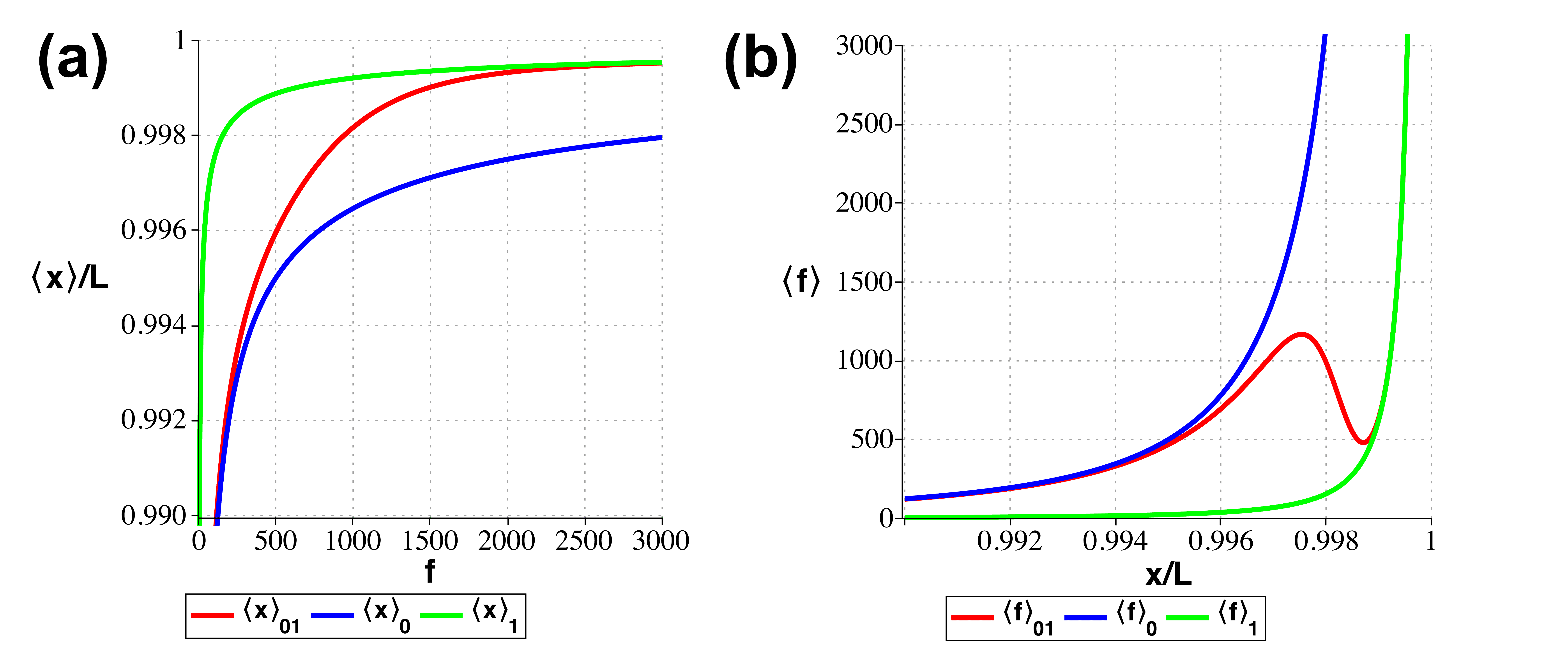}}
\caption{Force--extension response of a two-state polymer with reversible bending stiffness but no spontaneous curvature switching, \(c_0=0\). (a) Average extension in the Gibbs ensemble. (b) Average force in the Helmholtz ensemble. The red curve shows the two-state response, and the blue and green curves show the corresponding single-state branches with \(L_p^{(0)}=10\) and \(L_p^{(1)}=200\), respectively. Parameters: \(L=1\), \(\beta=1\), \(\epsilon=5\), \(c_0=0\), \(L_p^{(0)}=10\), and \(L_p^{(1)}=200\).
}
\label{fig:SP_b}
\end{figure}
\begin{figure*}[t]
\centering
\includegraphics[width=\textwidth]{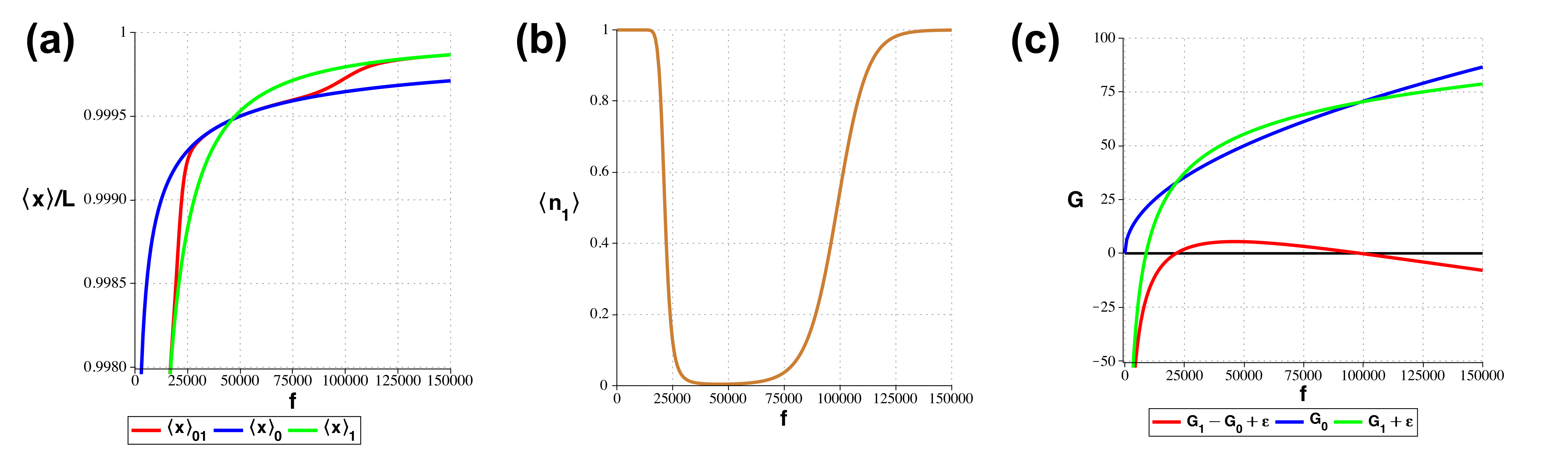}
\caption{Gibbs-ensemble response of a two-state stretched polymer in which both spontaneous curvature and bending stiffness change between states. (a) Mean extension \(\langle x\rangle/L\) as a function of the applied force. The red curve denotes the average response of the two-state chain, and the blue and green curves show the single-state responses of states \(0\) and \(1\), respectively. (b) Mean occupation \(\langle n_1\rangle_G\) of the curved/stiffer state. (c) Gibbs free energies of the two branches and their relative difference. The red curve represents \(\Delta G_{01}=G_1-G_0+\epsilon\), whose zero crossings indicate changes in the preferred state. The blue and green curves represent \(G_0\) and \(G_1+\epsilon\), respectively. \added{The second (reentrant) crossover occurs close to full extension, where the stiffness contrast dominates the relative Gibbs free energy.} Parameters: \(L=1\), \(\beta=1\), \(\epsilon=80\), \(c_0=2\), \(L_p^{(0)}=10\), and \(L_p^{(1)}=200\).}
\label{fig:SP_bc2_G}
\end{figure*}
\begin{figure*}[t]
\centering
\includegraphics[width=\textwidth]{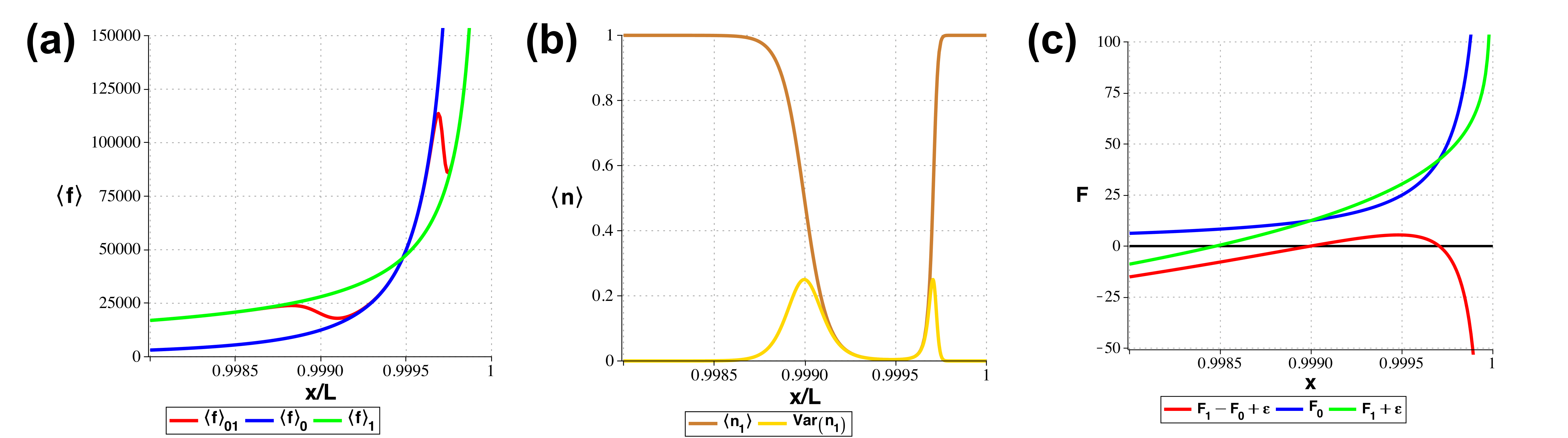}
\caption{Helmholtz-ensemble response of the same two-state polymer with simultaneous switching of spontaneous curvature and bending stiffness. (a) Mean force \(\langle f\rangle\) as a function of the normalized extension \(x/L\). The red curve is the two-state average response, while the blue and green curves are the single-state forces of states \(0\) and \(1\), respectively. The nonmonotonic portions of the red curve correspond to negative differential tensile stiffness. (b) Mean occupation \(\langle n_1\rangle_H\) of the curved state and the occupation variance \(\mathrm{Var}_H(n_1)=\langle n_1\rangle_H(1-\langle n_1\rangle_H)\). The peaks of \(\mathrm{Var}_H(n_1)\) mark the regions of strongest state mixing. (c) Helmholtz free energies of the two branches and their relative difference. The red curve shows \(\Delta F_{01}=F_1-F_0+\epsilon\), and its zero crossings correspond to the two state-switching events. The blue and green curves show \(F_0\) and \(F_1+\epsilon\), respectively. The second \added{(reentrant)} crossover occurs close to full extension, where the stiffness contrast dominates the relative Helmholtz free energy. Parameters: \(L=1\), \(\beta=1\), \(\epsilon=80\), \(c_0=2\), \(L_p^{(0)}=10\), and \(L_p^{(1)}=200\).}
\label{fig:SP_bc2_H}
\end{figure*}
\begin{figure}[h]
\centering{\includegraphics[width=0.5\textwidth]{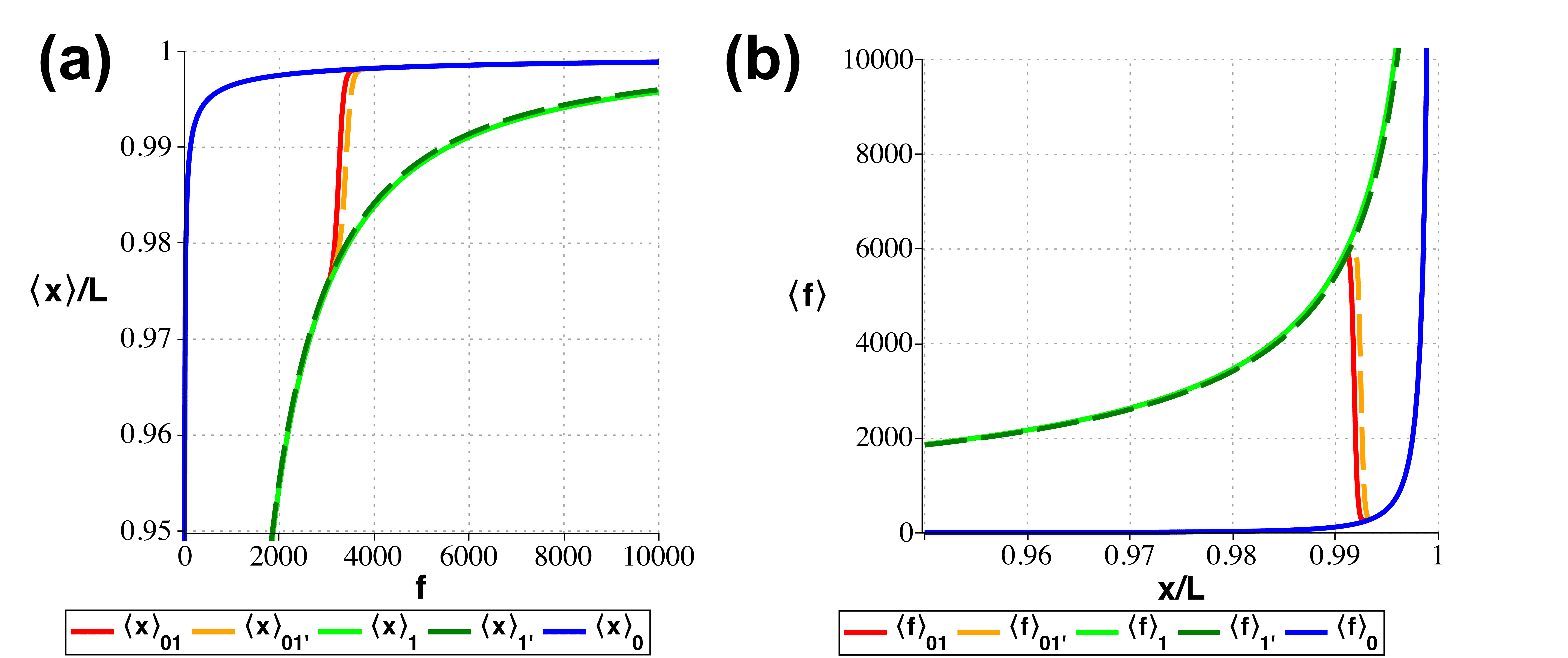}}
\caption{Effect of the curvature-dominated approximation on the two-state stretching response. (a) Gibbs-ensemble mean extension as a function of the applied force. (b) Helmholtz-ensemble mean force as a function of the normalized extension. The blue curves denote the uncurved state \(0\); the light-green and dashed green curves denote the full curved-state and approximate (athermal) branches, respectively; and the red and dashed orange curves denote the corresponding two-state averaged responses. Parameters: \(L=1\), \(\beta=1\), \(\epsilon=150\), \(c_0=2\), \(L_p^{(0)}=10\), \(L_p^{(1)}=200\).}
\label{fig:SP_aprx}
\end{figure}
In the numerical examples, we set \(c_0=2\), where \(c_0\) has the dimension of inverse length and represents the inverse radius of spontaneous curvature. For the parameters used here (\(L=1\)), this corresponds to \(c_0L=2\), indicating a relatively strongly curved unstretched configuration. Although such a curvature is close to the limit of the weak-bending approximation, the approximation remains appropriate in the strong-stretching regime, where the applied force suppresses transverse deflections and aligns the polymer nearly along the pulling direction. 

As we see in Figs.~\ref{fig:SP_c_G} and \ref{fig:SP_c_H}, for certain values of the control parameters, stretching induces a crossover from the curved $(1)$ to the uncurved $(0)$ state. Even though this crossover persists in both ensembles, the Gibbs and Helmholtz ensembles give significantly different elastic responses. Notably, in the Helmholtz ensemble, the crossover occurs in a region of negative differential extensibility. 

In real ligand-binding systems, ligand adsorption may affect not only the spontaneous curvature but also the bending stiffness of the polymer. In our analysis so far, which is illustrated by Figs.~\ref{fig:SP_c_G} and \ref{fig:SP_c_H}, we keep the bending stiffness fixed in order to isolate the effect of reversible spontaneous curvature. To assess how stiffness changes modify the response, we also consider two additional cases: reversible bending stiffness without spontaneous curvature switching (Fig.~\ref{fig:SP_b}), which was also analyzed in Ref.~\onlinecite{benetatos2022ensemble}, and simultaneous switching of bending stiffness and spontaneous curvature (Fig.~\ref{fig:SP_aprx}). \added{The relatively large values of \(\epsilon\) used in numerical examples are chosen to place the state crossovers within the displayed force or extension ranges and should be regarded as illustrative rather than system-specific.}

When the bending stiffness is allowed to fluctuate between the two states, we choose a large contrast in the persistence length. This estimate is motivated by the standard beam elasticity (see, e.g., Ref.~\onlinecite{howard2001mechanics}). For a slender filament with a circular cross section, the bending stiffness is \(\kappa = EI\), with \(I\propto R^4\). $E$ is the Young's modulus and $I$ is the second moment of inertia of the cross section. If a ligand with a size comparable to the polymer cross section attaches laterally, the effective second moment of inertia can increase substantially. Thus, \added{an order-of-magnitude} \deleted{a tenfold} change in the bending stiffness is a reasonable typical value to illustrate the effect of stiffness switching.


The behavior shown in the plots can be understood analytically as follows. In the Helmholtz ensemble, the force--extension curve can become nonmonotonic and exhibit negative differential tensile stiffness.
\begin{eqnarray}
\label{eq:diff_stiff_H}
\frac{\partial \langle f\rangle_{01}}{\partial x}
= \left(1-\langle n_1\rangle_H\right)\frac{\partial f_0}{\partial x}
+ \langle n_1\rangle_H\frac{\partial f_1}{\partial x} \nonumber\\
- \beta \mathrm{Var}_H\left(n_1\right) \left(f_1 - f_0\right)^2  
\end{eqnarray}
or equivalently,
\begin{eqnarray}
\frac{\partial^2 F_{01}}{\partial x^2}=\left(1-\langle n_1\rangle_H\right)\frac{\partial^2 F_0}{\partial x^2} 
+ \langle n_1\rangle_H\frac{\partial^2 F_1}{\partial x^2} \nonumber\\
- \beta \mathrm{Var}_H\left(n_1\right) \left(\frac{\partial F_1}{\partial x} - \frac{\partial F_0}{\partial x}\right)^2  
\end{eqnarray}
where 
\begin{eqnarray}
\mathrm{Var}\left(n_1\right) = \langle n_1^2\rangle - \langle n_1\rangle^2 \nonumber\\ 
=\langle n_1\rangle -\langle n_1\rangle^2 \nonumber\\
=\langle n_1\rangle \left(1-\langle n_1\rangle\right).  
\end{eqnarray}
There is a subtle relation between the state crossover and the negative differential stiffness. The crossover is controlled by the relative Helmholtz free energy \(\Delta F_{01}=F_1+\epsilon-F_0\), and occurs when \(\Delta F=0\), where the two states have equal statistical weight. A force mismatch between the two branches is not by itself sufficient to produce a negative slope, since away from the transition the occupation is nearly saturated and the two-state response simply follows one branch. The negative differential stiffness appears only in the switching region, where $\mathrm{Var}_H\left(n_1\right)$ emerges (Fig.~\ref{fig:SP_bc2_H}). Notice that the variance enters Eq.~\ref{eq:diff_stiff_H} with a minus sign. In this region, the redistribution of the statistical weight between branches with different forces gives a negative contribution to \(d\langle f\rangle_{01}/dx\), which can overcome the positive stiffness of single states.

By contrast, in the Gibbs ensemble, where the force is fixed, the contribution from the variance $\mathrm{Var}_G\left(n_1\right)$ is positive, and the average extension remains monotonic.
\begin{eqnarray}
\label{eq:diff_stiff_G}
\frac{\partial \langle x\rangle_{01}}{\partial f}
= - \frac{\partial^2 G_{01}}{\partial f^2}
= \left(1-\langle n_1\rangle_G\right)\frac{\partial x_0}{\partial f}
+ \langle n_1\rangle_G\frac{\partial x_1}{\partial f} \nonumber\\
+ \beta \mathrm{Var}_G\left(n_1\right) \left(x_1 - x_0\right)^2 \; >\; 0  
\end{eqnarray}
The physical effect of state switching differs in the two ensembles. In the Helmholtz ensemble, increasing the imposed extension favors the branch that requires the smaller force. In the Gibbs ensemble, by contrast, increasing the applied force favors the branch with the larger extension. This distinction is reflected in the thermodynamic relations of the two ensembles: in the Helmholtz ensemble \(f=\partial F/\partial x\), whereas in the Gibbs ensemble \(x=-\partial G/\partial f\). The opposite signs in these relations lead to opposite signs of the switching contribution. As a result, state switching reduces the differential stiffness in the Helmholtz ensemble, providing the origin of the negative extensibility. A similar negative extensibility in a two-state wormlike chain with reversible bending stiffness was discussed in Ref.~\onlinecite{benetatos2022ensemble}.

The \added{crossover} \deleted{transition} in the two-state response is an ensemble effect. The polymer does not mechanically interpolate between the two branches; rather, it occupies either state $0$ or state $1$, and the measured response is the probability-weighted average of the two single-state responses. The occupation probabilities are set by the relative thermodynamic potentials of the two states and by the activation energy. Because the theory assumes equilibrium sampling, experimental observation of this averaged transition requires switching between the states on the measurement time scale. Slow switching could lead to metastability or hysteresis instead of the equilibrium averaged curve.


The switching terms in Eq.~\ref{eq:diff_stiff_H} and Eq.~\ref{eq:diff_stiff_G} are the main source of ensemble inequivalence near a state transition. \added{They vanish} \deleted{It vanishes} when the occupation variance is small, i.e., when one state is almost fully occupied, and the response then reduces to that of a single mechanical branch. In this single-state limit, the Gibbs and Helmholtz responses are related by the Legendre transformation and ensemble equivalence is recovered. Thus, ensemble inequivalence is most pronounced in the state-mixing region, where both states are appreciably occupied and have distinct elasticity.

The Helmholtz response also reveals a secondary crossover when spontaneous curvature and bending stiffness change simultaneously. This behavior can be understood from the two competing tendencies. If only the bending stiffness changes, increasing the extension favors the branch with the larger persistence length, since its thermal stretching cost is smaller at a fixed \(x/L\). If only the spontaneous curvature changes, the curved state is more favored at smaller extensions. When both effects are present, the relative stability of the two states becomes nonmonotonic. The first change in occupation is controlled mainly by the curvature contribution, whereas sufficiently close to full extension the stiffness contrast dominates the Helmholtz free energy. As a result, the system can switch again to the stiffer curved state $n_1$, producing a second region of negative differential stiffness in the two-state force--extension curve, as shown in Fig.~\ref{fig:SP_bc2_H}. \deleted{The transitions happen when the $\langle n_1 \rangle_H$ as a function of $x$ curve has an inflection point (vanishing second derivative).} Because it is a sigmoidal curve for each transition, going from 1 to 0 and then back to 1, that happens when $\langle n_1 \rangle_H=1/2$. And that implies
\begin{eqnarray}
\Delta F_{01} = F_1-F_0+\epsilon=0.
\end{eqnarray}
If the two state average force has one transition, $\Delta F_{01}$ is monotonic.
At the local maximum of $\Delta F_{01}$, it satisfied 
\begin{eqnarray}
\frac{\partial \Delta F_{01}}{\partial x} = f_1-f_0=0 \;,
\end{eqnarray}
which means that the force in states 1 and 0 is the same. 
For unequal persistence lengths, this second crossover is expected in the high-extension limit and may therefore be observable only close to full extension of the chain. This reentrant crossover occurs not only in the Helmholtz, but also in the Gibbs ensemble. The main difference is that, in the Gibbs ensemble, the force-extension relation is always monotonic (as should be for reasons of thermodynamic stability~\cite{benetatos2022ensemble}). These results are illustrated in Figs.~\ref{fig:SP_bc2_G} and~\ref{fig:SP_bc2_H}.

It is also useful to compare the full curved-state description that includes thermal fluctuations with a simpler curvature-dominated approximation. The full strong-stretching response of a filament with constant spontaneous curvature contains both the athermal curvature contribution and the thermal fluctuation contribution, as seen in Eq.~\eqref{sp_const_Gc}. In the curvature-dominated approximation, we retain only the first correction in the curved branch,
\begin{eqnarray}
\langle x\rangle_{1'}(f)=L-L a f^{-3/2}\;,
\label{eq:curvature_dominated_branch}
\end{eqnarray}
while the uncurved state is represented by the thermal branch as in Eq.~\eqref{sp_const_Gth}. This approximation separates the curvature and thermal fluctuation contributions into two simple analytically tractable branches. As a result, the Gibbs and Helmholtz partition functions can be written in closed form, yielding explicit expressions for the mean extension, mean force, and curved-state occupation probability.

For the curvature-dominated branch, the corresponding Gibbs free energy is
\begin{eqnarray}
G_{1'}(f)=-fL-2La f^{-1/2}.
\end{eqnarray}
The relative Gibbs free \added{energy} \deleted{enregy} is therefore
\begin{eqnarray}
\Delta G_{01'}(f)
=G_{1'}(f)-G_0(f)+\epsilon \nonumber\\
=-2 La f^{-1/2} -2 L\added{b_0}\deleted{b} f^{1/2}+\epsilon .
\end{eqnarray}
The corresponding curved-state occupation probability is
\begin{eqnarray}
\langle n_{1'}\rangle_G(f)
=\frac{1}{\exp\left[\beta\Delta G_{01'}\right]+1}.
\end{eqnarray}
The two-state average extension in this approximation is
\begin{eqnarray}
\langle x\rangle_{01'}(f)
=\left(1-{\langle n_{1'} \rangle}_G\right) \langle x \rangle_0
+{\langle n_{1'} \rangle}_G \langle x \rangle_{1'}.
\end{eqnarray}

In the Helmholtz ensemble, defining \(\delta=1-x/L\), the Helmholtz free energy and force are
\begin{eqnarray}
F_{1'}(x)=-3L\left(a^2\delta\right)^{1/3}
\end{eqnarray}
and
\begin{eqnarray}
\langle f\rangle_{1'}(x)
= \left(a/\delta\right)^{2/3}\;,
\end{eqnarray}
respectively. The relative Helmholtz free energy is
\begin{eqnarray}
&\Delta F_{01'}(x)
=F_{1'}(x)+\epsilon-F_0(x) \nonumber\\
&=-3 L \left(a^2\delta\right)^{1/3}
- Lb_0^2\delta^{-1}+\epsilon . 
\end{eqnarray}
The corresponding curved-state occupation probability is
\begin{eqnarray}
\langle n_{1'}\rangle_H(x)
=\frac{1}{\exp\left[\beta\Delta F_{01'}\right]+1}.
\end{eqnarray}
Thus,
\begin{eqnarray}
\langle f\rangle_{01'}(x)
=\left(1-{\langle n_{1'} \rangle}_H\right)\langle f \rangle_0
+{\langle n_{1'} \rangle}_H\langle f \rangle_{1'} .
\end{eqnarray}

The validity of this approximation can be assessed by comparing the full curved response, Eq.~\eqref{sp_const_Gc}, with the curvature-dominated branch, Eq.~\eqref{eq:curvature_dominated_branch}. As shown in Fig.~\ref{fig:SP_aprx}, the two curves are nearly indistinguishable over the force range in which the curved state carries most of the statistical weight (weaker stretching). The difference between \(x_1\) and \(x_{1'}\) becomes relevant only after the system has already crossed over to the uncurved state. Therefore, for the purpose of describing the state selection and the onset of the curved state to uncurved state crossover, neglecting thermal fluctuations in the curvature-dominated branch provides a useful and transparent approximation.

Although this approximation is very good at capturing the first crossover from state 1 to state 0 (at a weaker stretching), it fails to reproduce the second (reentrant) crossover that appears when $L_p^{(1)}>L_p^{(0)}$. The reason is the neglect of the thermal fluctuations in state 1. On the other hand, it always yields an artificial secondary crossover at very strong stretching even for $L_p^{(1)}=L_p^{(0)}$. This artifact is not visible in the plots of Fig.~\ref{fig:SP_aprx}.

\section{\label{sec:SP_sin} Polymer under tension with reversible sinusoidal spontaneous curvature}

We next consider a curved state with a spatially modulated spontaneous curvature. In the weak-bending approximation, the governing equation is linear, and therefore a general curvature profile can be decomposed into Fourier modes. The sinusoidal profile studied here represents the simplest single-mode contribution and provides a natural extension of the constant-curvature case.

\begin{figure}[h]
\centering
\includegraphics[width=0.4\textwidth]{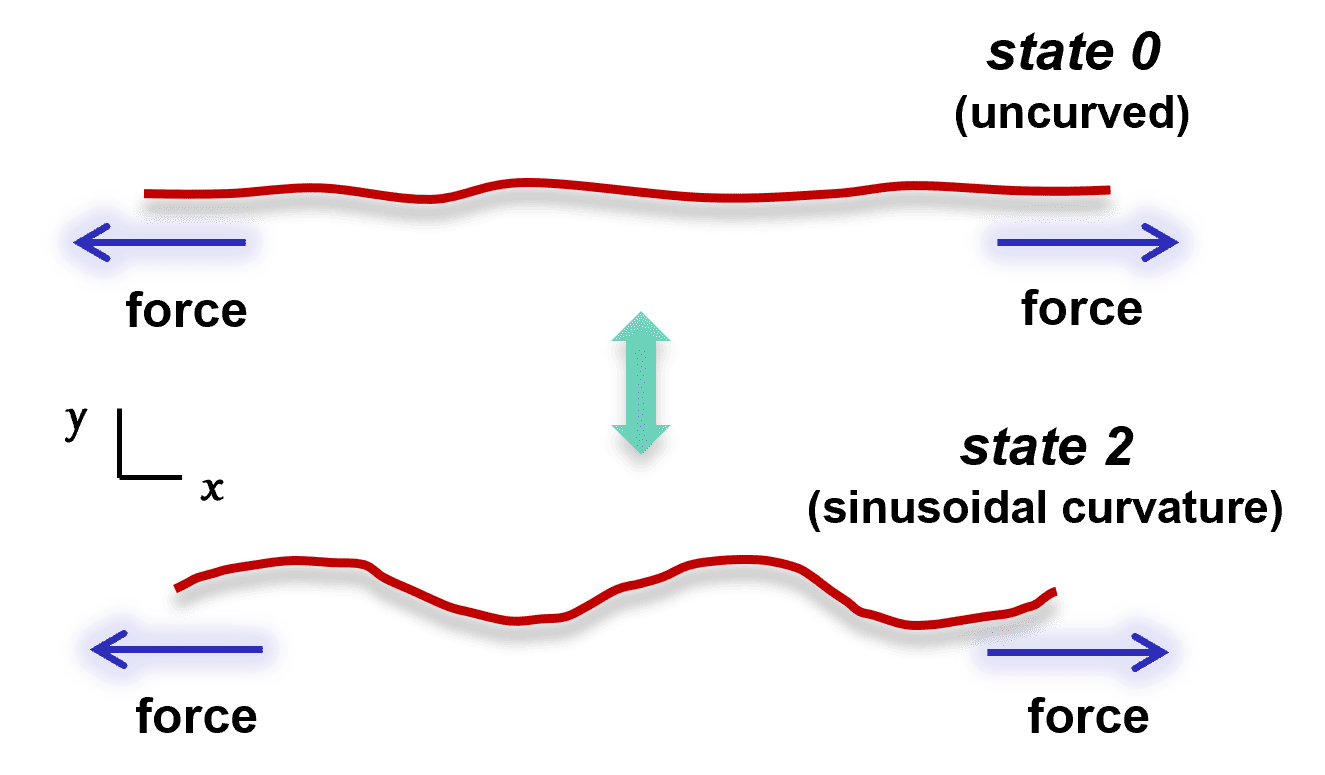}
\caption{Schematic of a stretched polymer reversibly switching between an uncurved thermal state and a curved state with sinusoidal spontaneous curvature.}
\label{fig:SP_sin_intro}
\end{figure}

We start from the strong-stretching force--extension relation for a filament with sinusoidal spontaneous curvature, obtained in Ref.~\onlinecite{benetatos2010stretching}.
\begin{eqnarray}
\label{sp_sin}
\frac{\langle x \rangle}{L} = 1 - 
\underbrace{\frac{1}{2}\frac{f_q f_c}{\left(f_q+f\right)^2}}_{\added{\mathrm{curvature\;contribution}}}
- \underbrace{\frac{1}{2}\left(\frac{f_{cr}}{f}\right)^{1/2}}_{\added{\mathrm{thermal\;fluctuation}}} \nonumber\\
= 1 - \frac{1}{2}\frac{f_q f_c}{\left(f_q+f\right)^2} - b f^{-1/2}\;. 
\end{eqnarray}
Here,
\begin{eqnarray}
f_{q} \equiv \kappa q^2
\end{eqnarray}
is the force scale associated with the wavelength of the sinusoidal curvature mode.

As in Sec.~\ref{sec:SP_const}, we use Eq.~\eqref{sp_sin} to construct a two-state description, schematically shown in Fig.~\ref{fig:SP_sin_intro}. State \(2\) denotes the curved state and contains both the sinusoidal-curvature contribution and the thermal fluctuation contribution. State \(0\) denotes the uncurved state, for which only the thermal fluctuation contribution remains. Thus, the extension of state \(2\) is
\begin{eqnarray}
\label{sp_sin_Gc}
\langle x \rangle_2(f)= L - \frac{L}{2}\frac{f_q f_c}{\left(f_q+f\right)^2}- Lb_2 f^{-1/2} .
\end{eqnarray}
Using \(\langle x\rangle=-\partial G/\partial f\), the Gibbs free energy of state \(2\) is
\begin{eqnarray}
G_2(f)
=-fL-\frac{L}{2}\frac{f_q f_c}{f_q+f}
+2Lb_2f^{1/2}.
\end{eqnarray}
Here, \(f_q\), \(f_c\), and \(b_2\) are all evaluated with the persistence length \(L_p^{(2)}\) of state \(2\).

The uncurved state \(0\) has the thermal WLC response
\begin{eqnarray}
\langle x \rangle_0(f) = L-Lb_0f^{-1/2},
\end{eqnarray}
with Gibbs free energy
\begin{eqnarray}
G_0(f)=-fL+2Lb_0f^{1/2}.
\end{eqnarray}
The Gibbs partition function is therefore
\begin{eqnarray}
Z_{02}^G(f)
=\exp\left[-\beta G_0\right]
+\exp\left[-\beta\left(G_2+\epsilon\right)\right],
\end{eqnarray}
where \(\beta=1/k_BT\). The probability of occupying the curved state in the Gibbs ensemble is
\begin{eqnarray}
\langle n_2 \rangle_G(f)
=\frac{1}{\exp\left[\beta\left(G_2-G_0+\epsilon\right)\right]+1}.
\end{eqnarray}
The average extension then follows as the occupation-weighted average of the two branch extensions:
\begin{eqnarray}
\langle x \rangle_{02}(f)
=\left(1-{\langle n_2 \rangle}_G\right)\langle x \rangle_0
+{\langle n_2 \rangle}_G\langle x \rangle_2 .
\end{eqnarray}

To formulate the Helmholtz description, we invert the force--extension relation of state \(2\). Defining \(\delta\equiv 1-x/L\), Eq.~\eqref{sp_sin_Gc} can be written as
\begin{eqnarray}
\delta=\frac{1}{2}\frac{f_qf_c}{(f_q+f)^2}+b_2 f^{-1/2}.
\label{eq:sp_sin_implicit_force}
\end{eqnarray}
For each fixed extension, the force of the curved branch, denoted by \(f_2(x)\), is obtained as the positive solution of Eq.~\eqref{eq:sp_sin_implicit_force}. This solution is unique for \(f>0\), since the right-hand side of Eq.~\eqref{eq:sp_sin_implicit_force} decreases monotonically with \(f\). Equivalently, the function 
\begin{eqnarray}
g(f)=\delta - \frac{1}{2}\frac{f_qf_c}{(f_q+f)^2}- b_2 f^{-1/2}
\end{eqnarray}
is monotonically increasing for \(f>0\). The explicit fifth-order polynomial obtained by setting \(t=\sqrt f\), together with the numerical root-finding procedure, is given in Appendix~\ref{appendix:sinusoidal_helmholtz}.

The Helmholtz free energy of state \(2\) is obtained from the Legendre transformation,
\begin{eqnarray}
F_2(x)=G_2(f_2)+f_2x .
\end{eqnarray}
Using Eq.~\eqref{sp_sin_Gc}, this gives
\begin{eqnarray}
F_2(x)=Lb_2\sqrt{f_2}
-\frac{L}{2}\frac{f_qf_c(f_q+2f_2)}{(f_q+f_2)^2}.
\label{eq:F2_sin}
\end{eqnarray}
The derivation of Eq.~\eqref{eq:F2_sin} is also included in Appendix~\ref{appendix:sinusoidal_helmholtz}.

For the uncurved state \(0\), the Helmholtz free energy and force are
\begin{eqnarray}
F_0(x) = Lb_0^2\left(1-\frac{x}{L}\right)^{-1},
\end{eqnarray}
and
\begin{eqnarray}
\langle f \rangle_0(x) = b_0^2\left(1-\frac{x}{L}\right)^{-2}.
\end{eqnarray}
The Helmholtz partition function is therefore
\begin{eqnarray}
Z_{02}^H(x)
=\exp\left[-\beta F_0\right]
+\exp\left[-\beta\left(F_2+\epsilon\right)\right].
\end{eqnarray}
The probability of occupying the curved state 2 is
\begin{eqnarray}
\langle n_2 \rangle_H(x)
=\frac{1}{\exp\left[\beta\left(F_2-F_0+\epsilon\right)\right]+1}.
\end{eqnarray}
The two-state Helmholtz free energy is
\begin{eqnarray}
F_{02}(x)=-k_BT\ln Z_{02}^H(x),
\end{eqnarray}
and the conjugate mean force follows from \(\langle f\rangle=\partial F/\partial x\). This gives the occupation-weighted branch average of the two single-state forces
\begin{eqnarray}
\langle f\rangle_{02}(x)
=\left(1-{\langle n_2 \rangle}_H\right)\langle f \rangle_0
+{\langle n_2 \rangle}_H\langle f \rangle_2.
\end{eqnarray}

\begin{figure}[h]
\centering{\includegraphics[width=0.5\textwidth]{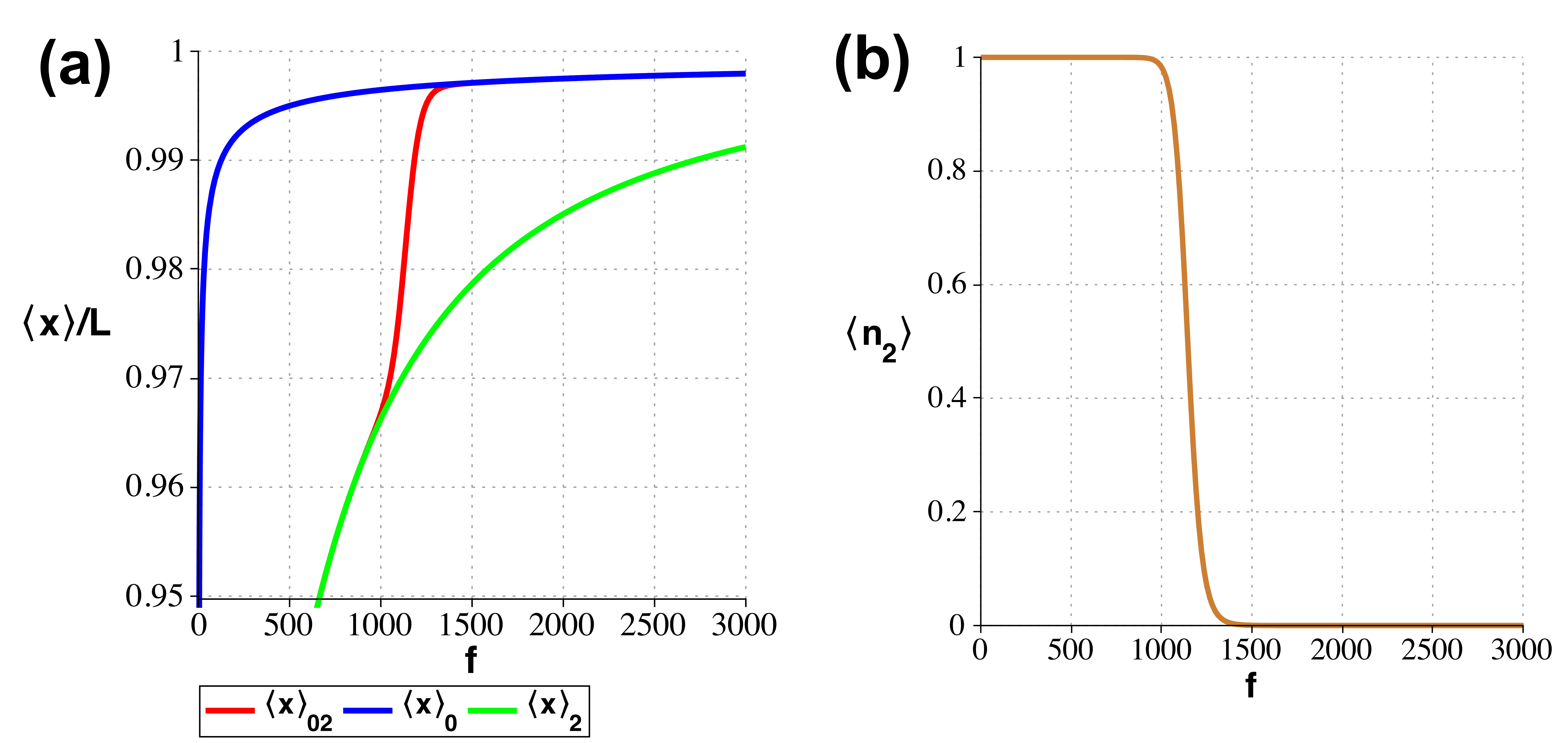}}
\caption{Analysis of the polymer in the Gibbs ensemble in sinusoidal case. (a) The red curve represents the average extension of the two-state system $\langle x \rangle_{02}$, while the blue and green curves correspond to the uncurved state $\langle x \rangle_0$ and the curved state $\langle x \rangle_2$, respectively. (b) Probability of the curved state, $\langle n_2 \rangle_G$, as a function of the applied force $f$. The calculation uses the following parameters: $L=1$, $\beta=1$, $\epsilon=50$, $c_0=7$, $L_p^{(0)}=L_p^{(2)}=10$, $q=4\pi$.}
\label{fig:SP_sin_G}
\end{figure}
\begin{figure}[h]
\centering{\includegraphics[width=0.5\textwidth]{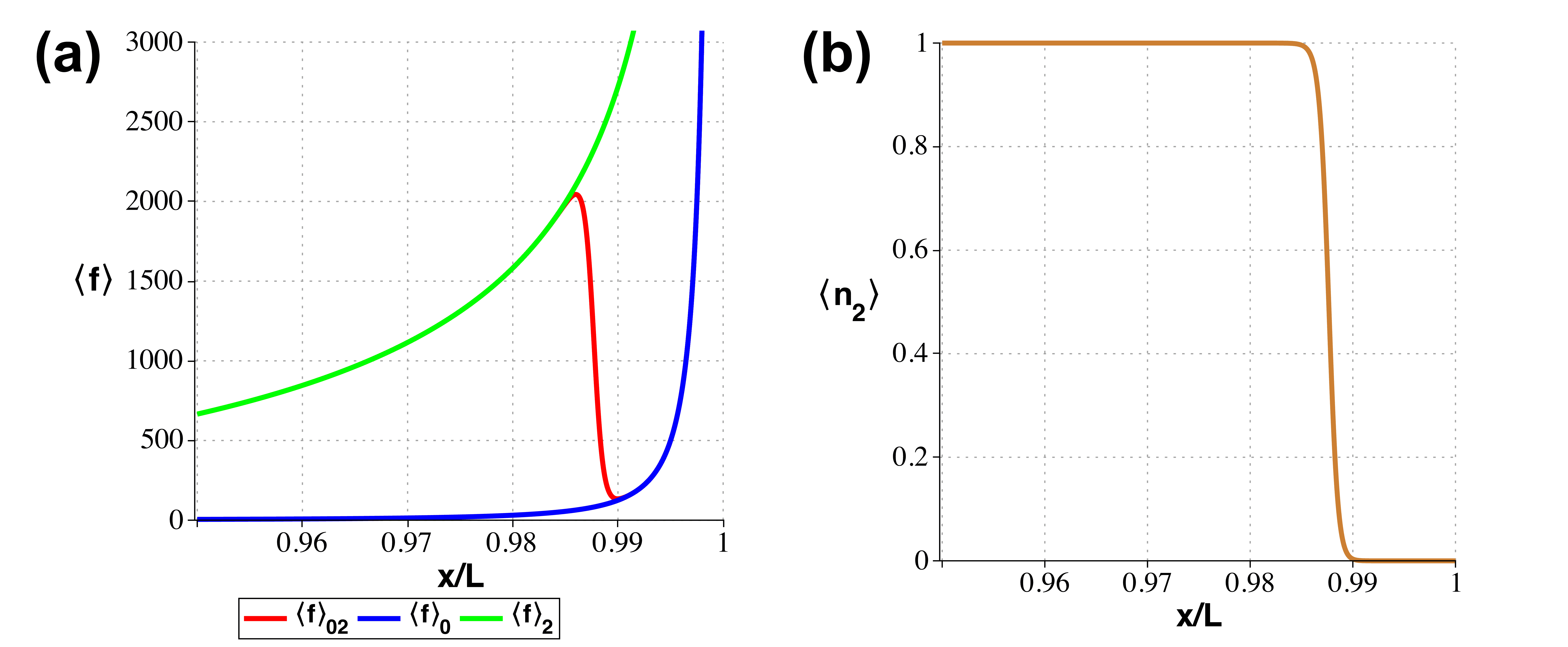}}
\caption{Analysis of the polymer in the Helmholtz ensemble in sinusoidal case. (a) The red curve indicates the average force $\langle f \rangle_{02}$ for the two-state system. The blue and green curves represent the forces for the uncurved state $\langle f \rangle_0$ and the curved state $\langle f \rangle_2$, respectively. (b) Probability of the curved state, $\langle n_2 \rangle_H$, plotted against the normalized extension $x/L$. The parameters are the same as in \added{Fig.~\ref{fig:SP_sin_G}} \deleted{Figure 3}: $L=1$, $\beta=1$, $\epsilon=50$, $c_0=7$, $L_p^{(0)}=L_p^{(2)}=10$, $q=4\pi$.}
\label{fig:SP_sin_H}
\end{figure}
\begin{figure*}[t]
\centering
\includegraphics[width=\textwidth]{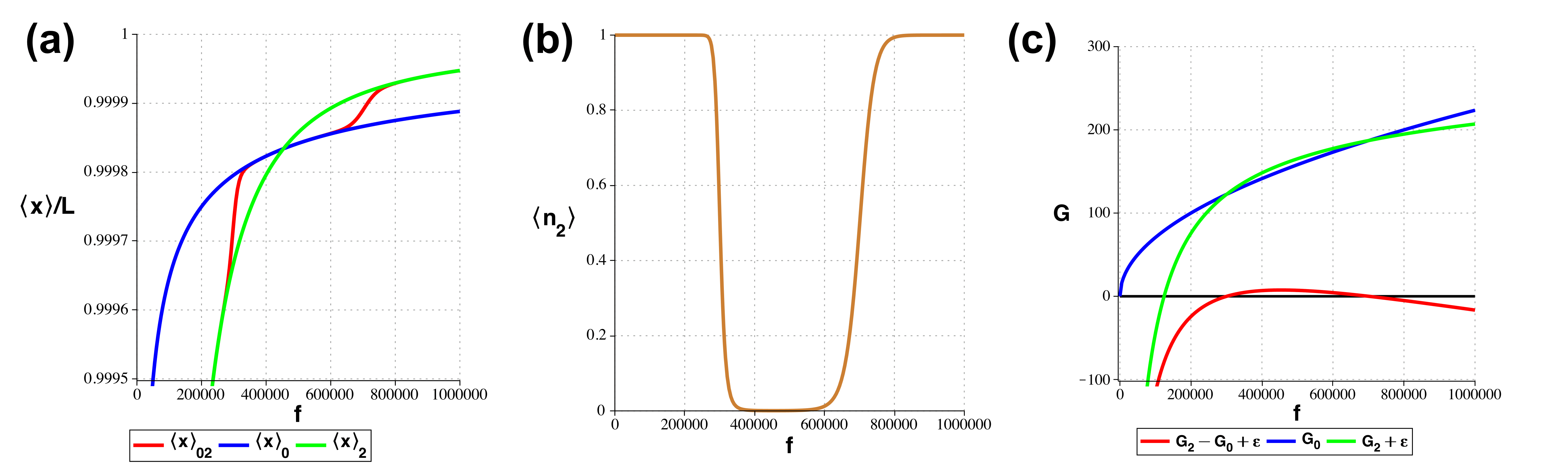}
\caption{Gibbs-ensemble response of a two-state stretched polymer in which both sinusoidal spontaneous curvature and bending stiffness change between states. (a) Mean extension \(\langle x\rangle/L\) as a function of the applied force. The red curve denotes the two-state average response, and the blue and green curves show the single-state responses of states \(0\) and \(2\), respectively. (b) Mean occupation \(\langle n_2\rangle_G\) of the curved state. (c) Gibbs free energies of the two branches and their relative difference. The red curve shows \(\Delta G_{02}=G_2-G_0+\epsilon\), whose zero crossings indicate changes in the preferred state. The blue and green curves show \(G_0\) and \(G_2+\epsilon\), respectively. \added{The second (reentrant) crossover occurs close to full extension, where the stiffness contrast dominates the relative Gibbs free energy.} Parameters: $L=1$, $\beta=1$, $\epsilon=185$, $c_0=6$, $L_p^{(0)}=10$, $L_p^{(2)}=200$, $q=4\pi$.}
\label{fig:SP_sin_bc2_G}
\end{figure*}
\begin{figure*}[t]
\centering
\includegraphics[width=\textwidth]{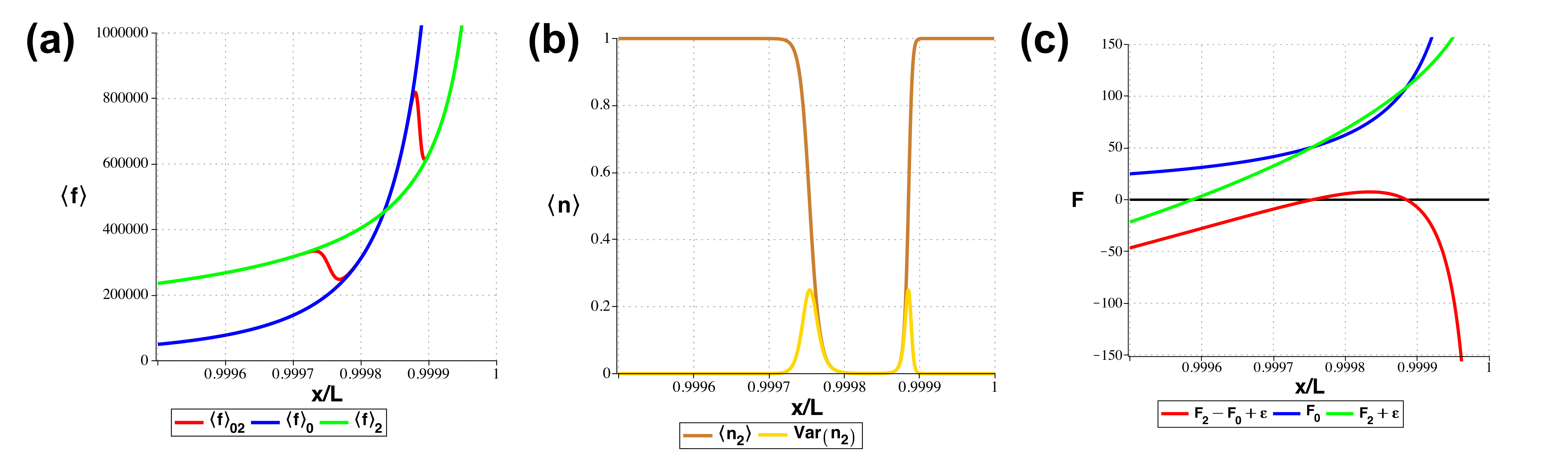}
\caption{Helmholtz-ensemble response of the same two-state polymer with simultaneous switching of sinusoidal spontaneous curvature and bending stiffness. (a) Mean force \(\langle f\rangle\) as a function of the normalized extension \(x/L\). The red curve is the two-state average response, while the blue and green curves are the single-state forces of states \(0\) and \(2\), respectively. The nonmonotonic portions of the red curve correspond to negative tensile differential stiffness. (b) Mean occupation \(\langle n_2\rangle_H\) of the curved state and the occupation variance \(\mathrm{Var}_H(n_2)=\langle n_2\rangle_H(1-\langle n_2\rangle_H)\). The peaks of \(\mathrm{Var}_H(n_2)\) mark the regions of strongest state mixing. (c) Helmholtz free energies of the two branches and their relative difference. The red curve shows \(\Delta F_{02}=F_2-F_0+\epsilon\), and its zero crossings correspond to the two state-switching events. The blue and green curves show \(F_0\) and \(F_2+\epsilon\), respectively. The second \added{(reentrant)} crossover occurs close to full extension, where the bending stiffness contrast dominates the relative Helmholtz free energy. Parameters: $L=1$, $\beta=1$, $\epsilon=185$, $c_0=6$, $L_p^{(0)}=10$, $L_p^{(2)}=200$, $q=4\pi$.}
\label{fig:SP_sin_bc2_H}
\end{figure*}
\begin{figure}[h]
\centering{\includegraphics[width=0.5\textwidth]{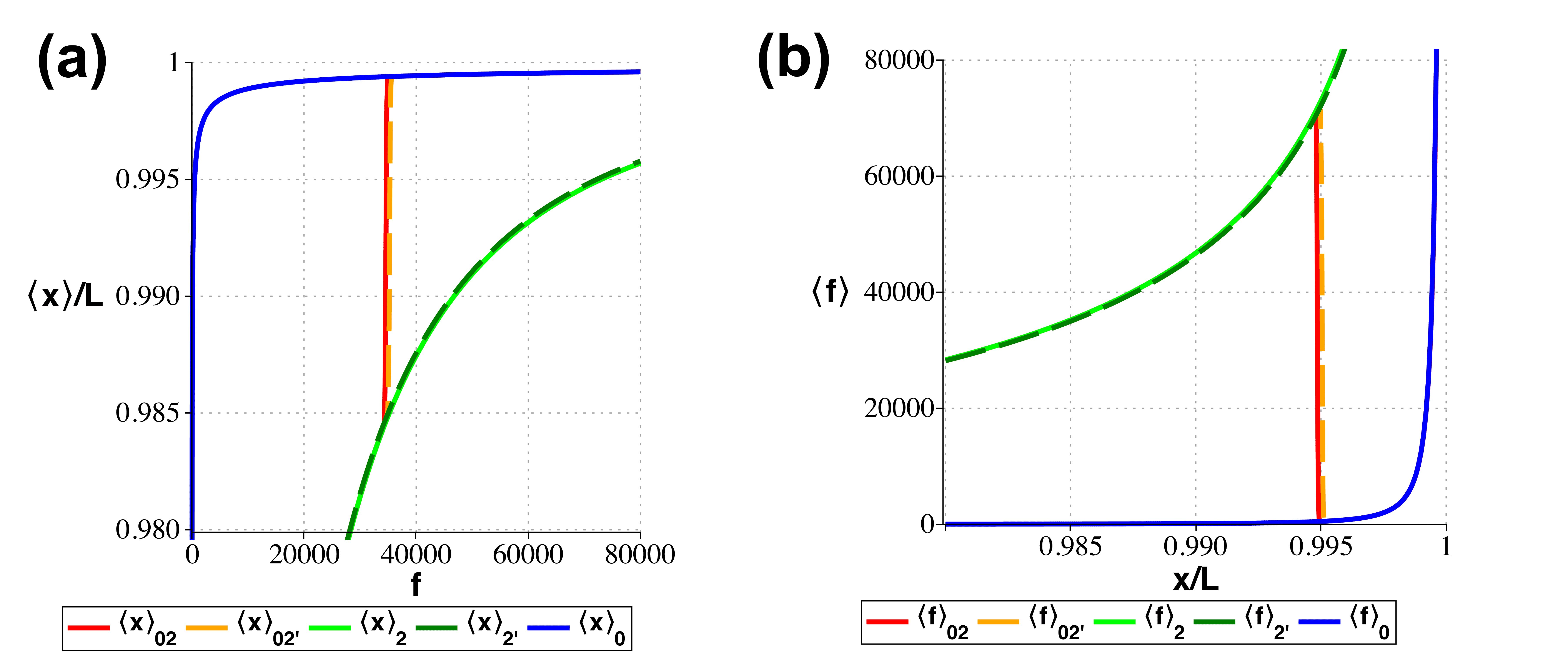}}
\caption{Effect of the curvature-dominated approximation on the two-state stretching response for a polymer with both reversible sinusoidal spontaneous curvature and reversible bending stiffness. (a) Gibbs-ensemble mean extension as a function of applied force. (b) Helmholtz-ensemble mean force as a function of normalized extension. The blue curves denote the uncurved state \(0\); the light-green and dashed green curves denote the full curved-state and approximate branches, respectively; and the red and dashed orange curves denote the corresponding two-state averaged responses. Parameters: $L=1$, $\beta=1$, $\epsilon=800$, $c_0=7$, $L_p^{(0)}=10$, $L_p^{(2)}=200$, $q=4\pi$.}
\label{fig:SP_aprx_2}
\end{figure}
The sinusoidal-curvature case exhibits the same qualitative behavior as the constant-curvature case. In the strong-stretching regime, the Gibbs and Helmholtz ensembles yield distinct elastic responses for the fluctuating two-state system. As common in two-state systems, the two averaged force--extension relations are inequivalent, and the corresponding free energies cannot be mapped onto each other by a simple Legendre transformation.

In the numerical examples, we use a larger curvature amplitude for the sinusoidal state than for the constant-curvature state. This choice is motivated by the zero-force weak-bending shape associated with a sinusoidal spontaneous curvature profile,
\begin{eqnarray}
y(s)=-\frac{c_0}{q^2}\sin(qs),
\end{eqnarray}
following Ref.~\onlinecite{benetatos2010stretching}. Thus, the transverse amplitude scales as \(c_0/q^2\). For \(q=4\pi\), the value \(c_0=2\) used in the constant-curvature case gives only a very small displacement amplitude, making the sinusoidal state nearly indistinguishable from the uncurved state. We therefore use \(c_0=7\) for the sinusoidal branch, while keeping \(q=4\pi\), so that the curvature modulation is large enough to produce a visible two-state response.

In Figs.~\ref{fig:SP_sin_G} and \ref{fig:SP_sin_H}, we see how the tensile response behaves in the Gibbs and Helmholtz ensemble, respectively, for a chain with reversible sinusoidal spontaneous curvature. In Figs.~\ref{fig:SP_sin_bc2_G} and \ref{fig:SP_sin_bc2_H}, we show the tensile response for a chain with simultaneously reversible bending stiffness and sinusoidal spontaneous curvature. As in the corresponding case with constant spontaneous curvature, we get a reentrant crossover to state 2 (spontaneous curvature and higher bending stiffness) at higher force/extension. 

As in Sec.~\ref{sec:SP_const}, we can capture the elastic response around the curved-uncurved crossover using an analytic approximation that neglects the thermal fluctuations of the curved state:
\begin{eqnarray}
\langle x \rangle_{2'}(f) = L - 
\frac{L}{2}\frac{f_q f_c}{\left(f_q+f\right)^2}.
\end{eqnarray}
For the curvature-dominated branch, the corresponding Gibbs free energy is
\begin{eqnarray}
G_{2'}(f)=-fL-\frac{L}{2}\frac{f_q f_c}{f_q+f}.
\end{eqnarray}
The relative Gibbs weight is therefore
\begin{eqnarray}
\Delta G_{02'}(f)
=G_{2'}(f)-G_0(f)+\epsilon \nonumber\\
=-\frac{L}{2}\frac{f_q f_c}{f_q+f}
-2 Lb_0 f^{1/2}+\epsilon .
\end{eqnarray}
The corresponding curved-state occupation probability is
\begin{eqnarray}
\langle n_{2'}\rangle_G(f)
=\frac{1}{\exp\left[\beta\Delta G_{02'}\right]+1}.
\end{eqnarray}
The two-state average extension in this approximation is
\begin{eqnarray}
\langle x\rangle_{02'}(f)
=\left(1-{\langle n_{2'} \rangle}_G\right)\langle x \rangle_0
+{\langle n_{2'} \rangle}_G\langle x \rangle_{2'} .
\end{eqnarray}

In the Helmholtz ensemble, defining \(\delta=1-x/L\), the Helmholtz free energy and force are
\begin{eqnarray}
F_{2'}(x)
= -2L\left(\frac{f_q f_c}{2}\delta\right)^{1/2} + L f_q \delta
\end{eqnarray}
and
\begin{eqnarray}
\langle f \rangle_{2'}(x)
= \left(\frac{2 }{f_q f_c}\delta\right)^{-1/2} - f_q .
\end{eqnarray}
The relative Helmholtz weight is
\begin{eqnarray}
\Delta F_{02'}(x)
=F_{2'}(x)-F_0(x)+\epsilon \nonumber\\
= -2 L\left(\frac{f_q f_c}{2}\delta\right)^{1/2} + L f_q \delta \nonumber\\
- Lb_0^2\delta^{-1}+\epsilon . 
\end{eqnarray}
The corresponding curved-state occupation probability is
\begin{eqnarray}
\langle n_{2'}\rangle_H(x)
=\frac{1}{\exp\left[\beta\Delta F_{02'}\right]+1}.
\end{eqnarray}
Thus,
\begin{eqnarray}
\langle f\rangle_{02'}(x)
=\left(1-{\langle n_{2'} \rangle}_H\right)\langle f \rangle_0
+{\langle n_{2'} \rangle}_H\langle f \rangle_{2'} .
\end{eqnarray}
This approximation is illustrated in Fig.~\ref{fig:SP_aprx_2}. As in the corresponding case with constant spontaneous curvature, the same caveats apply. 


The sinusoidal-curvature case therefore supports the same central conclusion as the constant-curvature case. Reversible spontaneous curvature produces ensemble-dependent elasticity because the occupation of the curved state is controlled by the relative thermodynamic potential of the two internal states. The spatial modulation of the curvature changes the quantitative location and sharpness of the crossover, through the force scale \(f_q\), but it does not change the underlying mechanism. When the bending stiffness is fixed, stretching suppresses the curved state and shifts the statistical weight toward the uncurved thermal state. When the bending stiffness is also reversible, the competition between curvature suppression at intermediate extension and the high-extension preference for the stiffer state can again produce a reentrant crossover. Thus, the ensemble inequivalence, the negative differential stiffness, and the reentrant crossover found in this work are not special consequences of a uniform spontaneous curvature. Rather, they arise more generally from the coupling between mechanical deformation and the force- or extension-dependent occupation of internal curvature states (which may also differ in bending stiffness). A multi-state extension of this free-energy description is discussed in Appendix~\ref{appendix:three_state}.

\section{\label{sec:TA}Grafted Polymer Subjected to a Bending Torque}

In this Section, we consider a two-state semiflexible polymer grafted to a rigid wall and subjected to a bending torque at its free end (in two dimensions). We have also examined the case of a grafted polymer under a transverse bending force \added{in Sec.~\ref{sec:FY}.} \deleted{; however, that system does not show ensemble inequivalence within the present formulation (weakly bending approximation). It is therefore discussed separately in Appendix~\ref{appendix:bending_force}.}

A possible realization of the torque-controlled setup is a magnetic bead attached to the free end of the polymer. We assume that the bead carries a magnetic dipole moment \(\vec{\mu}\), which remains aligned with the tangent direction at the polymer tip. When an external magnetic field \(\vec{B}\) is applied along the \(+y\) direction, the bead experiences a torque that tends to rotate the tip angle \(\theta\). We define the effective torque amplitude as \(\tau \equiv \mu B\). In the small-angle limit, the magnetic coupling is written as
\begin{eqnarray}
E_m=&-\vec{\mu} \cdot \vec{B} =- \mu B \cos\left(\frac{\pi}{2}-\theta\right)\nonumber\\
=&-\mu B \sin\theta \approx -\mu B \theta =- \tau \theta
\end{eqnarray}
With the sign convention used below, this coupling enters the effective Hamiltonian through the term \(-\tau\theta\).

\begin{figure}[h]
\centering
\includegraphics[width=0.3\textwidth]{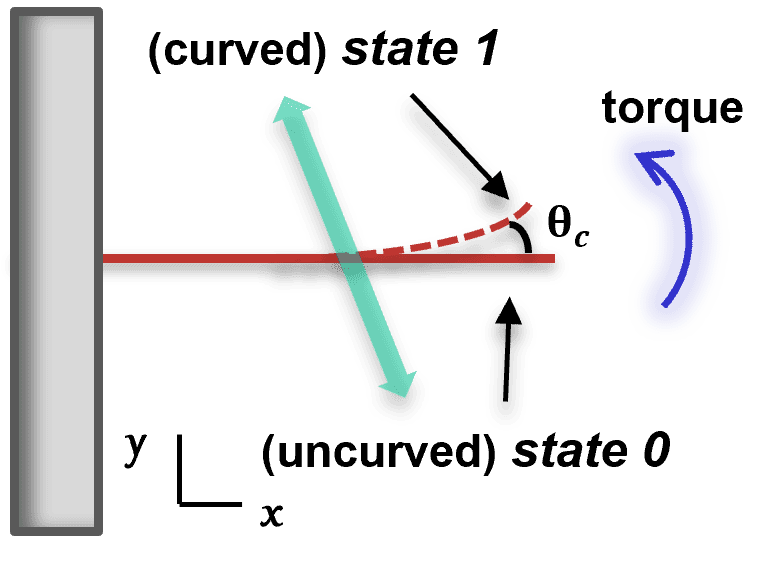}
\caption{Schematic of a grafted two-state semiflexible polymer under a bending torque.}
\label{fig:TA_intro}
\end{figure}

We model the polymer as a two-state system. The state \(n=0\) represents an uncurved branch, whereas \(n=1\) represents a curved branch with spontaneous tip angle \(\theta_c\). For a given internal state \(n\) and tip angle \(\theta\), the coarse-grained effective Hamiltonian is
\begin{eqnarray}
H_n=\frac{1}{2}\kappa_n(\theta-n\theta_c)^2-\tau\theta+n\epsilon .
\end{eqnarray}

\begin{eqnarray}
\nonumber
n =
\begin{cases}
0 \quad\text{(uncurved)} \\
1 \quad\text{(curved)}
\end{cases}
\end{eqnarray}
Here, \(\kappa_n\) is the effective angular bending stiffness (units of energy) of state \(n\), and \(\epsilon\) is the activation energy associated with occupying the curved state. It is related to the bending stiffness of the grafted wormlike chain $\tilde{\kappa}$ and its total contour length $L$ via $\kappa=\tilde{\kappa}/\added{L}\deleted{2L}$~\cite{Razbin_PB_magnetic_nunchucks,benetatos2021orientational}. The final two terms describe, respectively, the coupling to the applied torque and the energetic cost of the curved state (Figure~\ref{fig:TA_intro}).

In the Gibbs ensemble, the applied magnetic field, and hence the torque \(\tau\), is externally controlled, while the tip angle \(\theta\) is allowed to fluctuate thermally. The partition function is obtained by integrating over all possible tip angles~\cite{Razbin_PB_magnetic_nunchucks,benetatos2021orientational} and summing over the two internal states as
\begin{eqnarray}
Z_{01}^G(\tau)=\int_{-\infty}^{+\infty}d\theta\sum_{n=0}^{1}\exp\left[-\beta H_n\right]
\end{eqnarray}
where $\beta=1/k_BT$.
Performing the Gaussian integration gives
\begin{eqnarray}
Z_{01}^G(\tau)=\exp[-\beta\Phi_0^G]+\exp[-\beta\Phi_1^G]
\end{eqnarray}
where
\begin{eqnarray}
\Phi_0^G(\tau)
=-\frac{\tau^2}{2\kappa_0}
+\frac{1}{2\beta}\ln\left(\frac{\beta\kappa_0}{2\pi}\right)
\end{eqnarray}
and
\begin{eqnarray}
\Phi_1^G(\tau)
=-\frac{\tau^2}{2\kappa_1}
-\tau\theta_c+\epsilon
+\frac{1}{2\beta}\ln\left(\frac{\beta\kappa_1}{2\pi}\right).
\end{eqnarray}
These partition functions determine the occupation probability of the curved state. In the Gibbs ensemble, this probability is
\begin{eqnarray}
{\langle n_1 \rangle}_G(\tau)=
\frac{1}{\exp\left[{\beta\left(\Phi_1^G-\Phi_0^G\right)}\right]+1} .
\end{eqnarray}
The mean tip angle is obtained as the ensemble average over both the continuous variable \(\theta\) and the discrete state variable \(n\).
\begin{eqnarray}
\langle \theta \rangle_{01}(\tau)=\frac{\int_{-\infty}^{+\infty}d\theta\sum_{n=0}^{1}\theta\exp\left[-\beta H_n\right]}
{\int_{-\infty}^{+\infty}d\theta\sum_{n=0}^{1}\exp\left[-\beta H_n\right]}
\end{eqnarray}
The denominator is the Gibbs partition function \(Z_{01}^G\). The numerator 
can also be evaluated by Gaussian integration:
\begin{eqnarray}
\left(\frac{\tau}{\kappa_0}\right) \exp[-\beta\Phi_0^G] +\left(\frac{\kappa_1\theta_c+\tau}{\kappa_1}\right)\exp[-\beta\Phi_1^G]
\end{eqnarray}
Substituting this expression into the ratio above, we obtain the torque-dependent mean tip angle as
\begin{eqnarray}
\langle \theta \rangle_{01}(\tau)
=\left(1 - {\langle n_1 \rangle}_G\right)\langle\theta\rangle_0 + {\langle n_1 \rangle}_G \langle\theta\rangle_1.
\end{eqnarray}
The torque--angle relations of the two individual branches are 
\begin{eqnarray}
\langle\theta\rangle_0(\tau)=\frac{\tau}{\kappa_0}, \qquad 
\langle\theta\rangle_1(\tau)=\frac{\kappa_1\theta_c+\tau}{\kappa_1}.
\end{eqnarray}

In the Helmholtz ensemble, the tip angle \(\theta\) is fixed, and the system fluctuates between the two internal states. The corresponding partition function is therefore a discrete sum over \(n=0\) and \(n=1\),
\begin{eqnarray}
Z_{01}^H(\theta)=\exp[-\beta\Phi_0^H]+\exp[-\beta\Phi_1^H]\;,
\end{eqnarray}
where
\begin{eqnarray}
\label{eq:TA_H0}
\Phi_0^H(\theta)
=\frac{1}{2}\kappa_0\theta^2
\end{eqnarray}
and
\begin{eqnarray}
\label{eq:TA_H1}
\Phi_1^H(\theta)
=\frac{1}{2}\kappa_1(\theta-\theta_c)^2 + \epsilon .
\end{eqnarray}
\added{Note that Eqs.~\eqref{eq:TA_H0} and \eqref{eq:TA_H1} without $\epsilon$, are consistent with the fixed-angle free energies that can get from the joint probability density of Eq.~\eqref{eq:bf_joint_distribution} after integrating out $x$ and $y$.} In the Helmholtz ensemble, the occupation probability of the curved state is given by
\begin{eqnarray}
\langle n_1 \rangle_H(\theta)
=\frac{1}{\exp\left[{\beta\left(\Phi_1^H-\Phi_0^H\right)}\right]+1} .
\end{eqnarray}
The Helmholtz free energy is obtained by \(F=-k_BT\ln{Z^H}\). The average conjugate torque that corresponds to a fixed tip angle is then obtained from \(\langle \tau \rangle(\theta)=\partial F/\partial \theta\).
\begin{eqnarray}
\langle \tau \rangle_{01}(\theta)
=\left(1 - {\langle n_1 \rangle}_H\right)\langle \tau \rangle_0
+ {\langle n_1 \rangle}_H \langle \tau \rangle_1
\end{eqnarray}
The torque--angle relations of the two individual branches are
\begin{eqnarray}
\langle \tau \rangle_0(\theta) = \kappa_0 \theta, \qquad 
\langle \tau \rangle_1(\theta) = \kappa_1 \left( \theta - \theta_c\right).
\end{eqnarray}

\begin{figure}[h]
\centering{\includegraphics[width=0.5\textwidth]{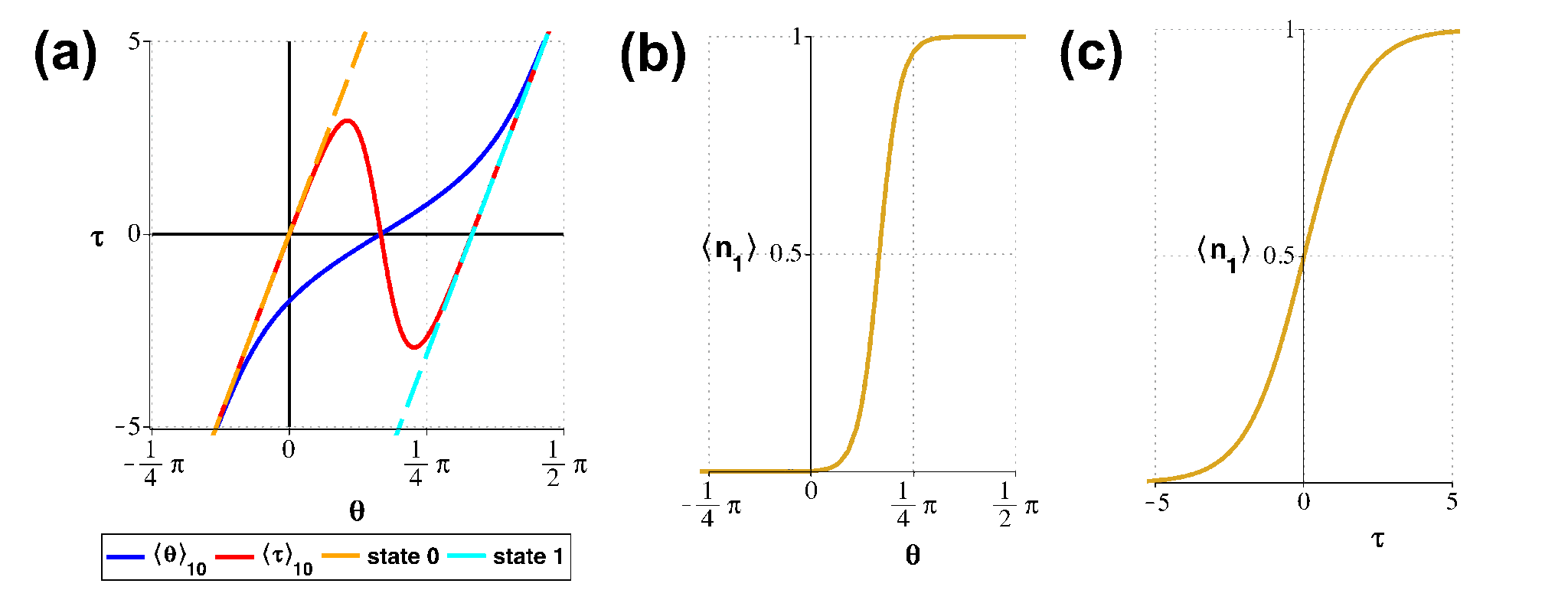}}
\caption{Bending response of a grafted rodlike polymer with reversible curvature  in the Helmholtz and Gibbs ensembles. (a) Torque–angle curves. The red and blue curves indicate the average torque $\langle \tau \rangle_{01}$ and the average angle $\langle \theta \rangle_{01}$ for the two-state system, respectively. The dashed orange and cyan lines represent the forces for the uncurved state 0 and the curved state 1, respectively. (b) Probability of the curved state, $\langle n_1 \rangle_H$, plotted against the normalized angle $\theta$. (c) Probability of the curved state, $\langle n_1 \rangle_G$, plotted against the normalized torque $\tau$. The calculation uses the following parameters: $\beta=1$, $\epsilon=0$, $\theta_c=\pi/3$, $\kappa_0=\kappa_1=12$.}
\label{fig:TA_c}
\end{figure}
\begin{figure}[h]
\centering{\includegraphics[width=0.5\textwidth]{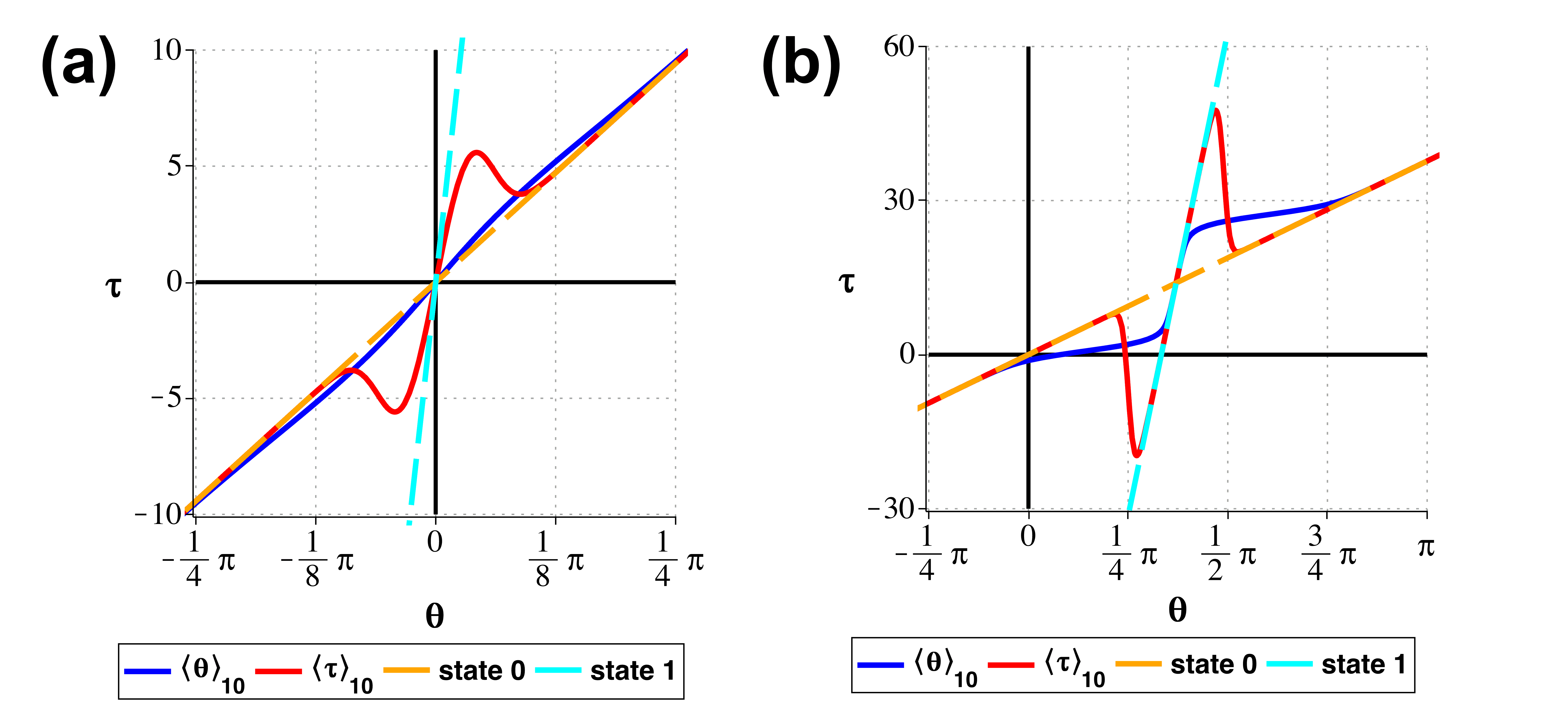}}
\caption{Torque--angle responses for two-state grafted polymers. (a) Bending-stiffness switching only, with \(\theta_c=0\). (b) Simultaneous switching of bending stiffness and spontaneous angle, with \(\theta_c=\pi/3\). The red curves denote the Helmholtz response \(\langle \tau\rangle_{01}\), and the blue curves denote the Gibbs response \(\langle \theta\rangle_{01}\) plotted in the torque--angle plane. The orange and cyan dashed lines are the single-state branches for states \(0\) and \(1\), respectively. Parameters: \(\beta=1\), \(\epsilon=0\), \(\kappa_0=12\), and \(\kappa_1=120\).
}
\label{fig:TA_b_bc}
\end{figure}
Figure~\ref{fig:TA_c} shows that the Gibbs and Helmholtz ensembles give distinct torque--angle responses for this finite two-state system. The sign and magnitude of the applied torque bias the occupation of the two states: a torque that rotates the polymer toward the spontaneous angle favors the curved state, whereas a torque in the opposite direction suppresses it. In Fig.~\ref{fig:TA_b_bc}, we show the torque-angle response for the case where we have switching in the bending stiffness only (a) and for the case where we have simultaneous switching of bending stiffness and spontaneous curvature (b). \added{When spontaneous curvature and bending stiffness switch simultaneously, their competition can produce reentrant behavior in the torque--angle response. A torque applied in the direction of the preferred curvature initially favors the curved state, whereas sufficiently strong bending favors the softer state because it has a larger angular compliance. This differs from stretching, where a strong tensile force straightens the polymer and can favor the stiffer state near full extension. Consequently, the system can switch from the softer state to the curved, stiffer state and then back to the softer state as the torque or imposed angle is increased.} It is worth pointing out that the response in the Gibbs ensemble is always monotonic, as dictated by thermodynamic stability~\cite{benetatos2022ensemble}. We should also clarify that the range of angles shown in the plots seems to exceed the limits of the weakly bending approximation. However, we do not expect any qualitative change in the behavior of the polymer in this range. 





The magnetic bead construction introduced above provides a natural physical realization of the Gibbs ensemble, since the external magnetic field controls the applied torque. By contrast, the Helmholtz ensemble would require direct control of the tip angle. Such a fixed-angle condition could in principle be approached through mechanical clamping or angular confinement in an ordered medium({\it e.g.,} nematic liquid crystal), but its experimental realization is less direct. We therefore treat the Helmholtz ensemble primarily as the fixed-angle thermodynamic counterpart of the torque-controlled setup.

\section{\label{sec:FY} Grafted Polymer Subjected to a Bending Force}

\begin{figure}[h]
\centering{\includegraphics[width=0.3\textwidth]{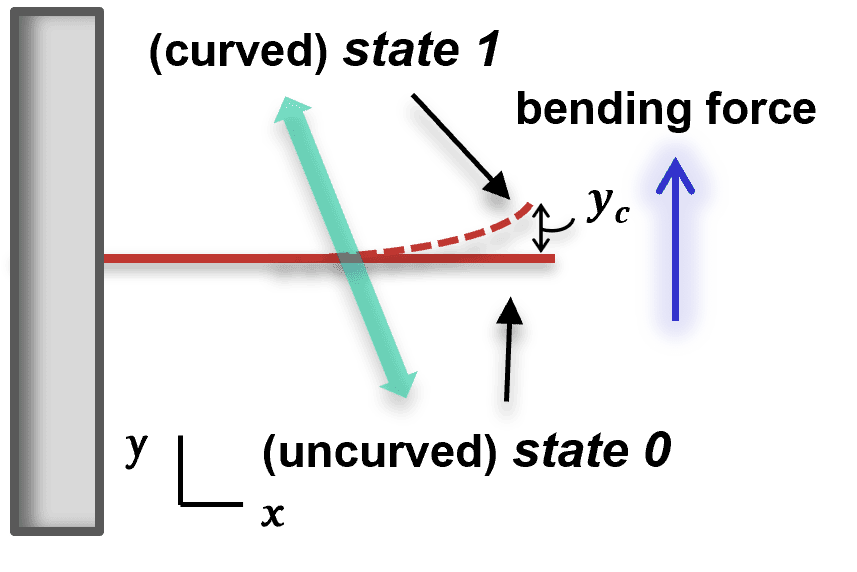}}
\caption{Schematic of a grafted two-state semiflexible polymer under a bending force.}
\label{fig:FY_intro}
\end{figure}

In this section, we consider a two-state semiflexible polymer grafted to a rigid wall and subjected to a transverse bending force at its free end. This geometry extends the preceding stretching analysis to a grafted setup in which the controlled variables are the physical transverse tip displacement \(y\) in the Helmholtz ensemble and the conjugate transverse force \(f_y\) in the Gibbs ensemble. For a filament that can reversibly acquire spontaneous curvature, the curved state has a shifted preferred configuration relative to the uncurved state. We use this weak-bending formulation to examine how internal state switching affects the transverse force--displacement response, as shown in Fig.~\ref{fig:FY_intro}. 

The Helmholtz ensemble corresponds to fixed transverse displacement \(y\). Integrating out the angular degree of freedom from Eq.~\eqref{eq:bf_joint_distribution}, we obtain
\begin{eqnarray}
Z_1^{H}(y)
=\int_{-\infty}^{\infty}d\theta\,Z(\theta,x,y;s) = \exp\left[-\beta F_1 \right] ,
\end{eqnarray}
where
\begin{eqnarray}
F_1 (y) = \frac{K_1}{2}\left(y-y_c\right)^2 +\frac{1}{2\beta}\ln{\left(\frac{2\pi}{\beta K_1}\right)} .
\end{eqnarray}
Here,
\begin{eqnarray}
K_i(s)=\frac{3L_p^{(i)}}{2\beta s^3}, \qquad i\in\{0,1\}.
\end{eqnarray}
The Helmholtz free energy of the uncurved state is obtained by setting $y_c=0$.
\begin{eqnarray}
F_0 (y) = \frac{K_0}{2}y^2 +\frac{1}{2\beta}\ln{\left(\frac{2\pi}{\beta K_0}\right)} .
\end{eqnarray}
The conjugate forces are
\begin{eqnarray}
\langle f_y\rangle_0(y) =K_0y ,\qquad
\langle f_y\rangle_1(y) =K_1 \left(y-y_c\right) .
\end{eqnarray}
In the shifted coordinate \(\tilde y=y-y_c\), the Gaussian kernel has the same form as that of an uncurved grafted filament. However, in terms of the physical lab-frame transverse displacement \(y\), the distribution is centered at \(y_c\). Thus, spontaneous curvature shifts the preferred transverse displacement of the curved state and therefore affects the two-state force--displacement response.

In the Gibbs ensemble, the transverse force \(f_y\) is fixed and couples linearly to the tip displacement through the work term \(-f_y y\). The Gibbs partition function is
\begin{eqnarray}
Z_1^{G}(f_y)
=\int_{-\infty}^{\infty}dy\,
Z_1^{H}(y)\exp(\beta f_y y) = \exp\left[-\beta G_1 \right].
\end{eqnarray}
where
\begin{eqnarray}
G_1(f_y)=-\frac{\beta s^3}{3L_p^{(1)}}f_y^2 -y_c f_y .
\end{eqnarray}
The Gibbs free energy of the uncurved state is
\begin{eqnarray}
G_0(f_y)=-\frac{\beta s^3}{3L_p^{(0)}} f_y^2 .
\end{eqnarray}
The conjugate displacements are
\begin{eqnarray}
\langle y \rangle_0(f_y)=\frac{f_y}{K_0}, \quad
\langle y \rangle_1(f_y)=\frac{f_y}{K_1} +y_c .
\end{eqnarray}
We now extend this result to a two-state curvature model. The Helmholtz partition function of the two-state system is
\begin{eqnarray}
Z_{01}^H(y)=Z_0^{H}(y)+e^{-\beta\epsilon}Z_1^{H}(y) \nonumber\\
=\exp\left[-\beta F_0\right]
+\exp\left[-\beta\left(F_1+\epsilon\right)\right].
\end{eqnarray}
The occupation probability of the curved state is
\begin{eqnarray}
\langle n_1\rangle_H
=\frac{1}{\exp\left[\beta\left(F_1-F_0+\epsilon\right)\right]+1}.
\end{eqnarray}
Therefore,
\begin{eqnarray}
\langle f_y\rangle_{01}(y)
=\left(1-{\langle n_{1} \rangle}_H\right)\langle f_y \rangle_0
+{\langle n_{1} \rangle}_H\langle f_y \rangle_{1} .
\end{eqnarray}
Similarly, in the Gibbs ensemble,
\begin{eqnarray}
Z_{01}^G(f_y)
=\exp\left[-\beta G_0\right]
+\exp\left[-\beta\left(G_1+\epsilon\right)\right].
\end{eqnarray}
The occupation probability of the curved state is
\begin{eqnarray}
\langle n_1\rangle_G
=\frac{1}{\exp\left[\beta\left(G_1-G_0+\epsilon\right)\right]+1}.
\end{eqnarray}
Therefore,
\begin{eqnarray}
\langle y\rangle_{01}(f_y)=\left(1-{\langle n_{1} \rangle}_G\right)\langle y \rangle_0
+{\langle n_{1} \rangle}_G\langle y \rangle_{1} .
\end{eqnarray}

\begin{figure}[h]
\centering{\includegraphics[width=0.5\textwidth]{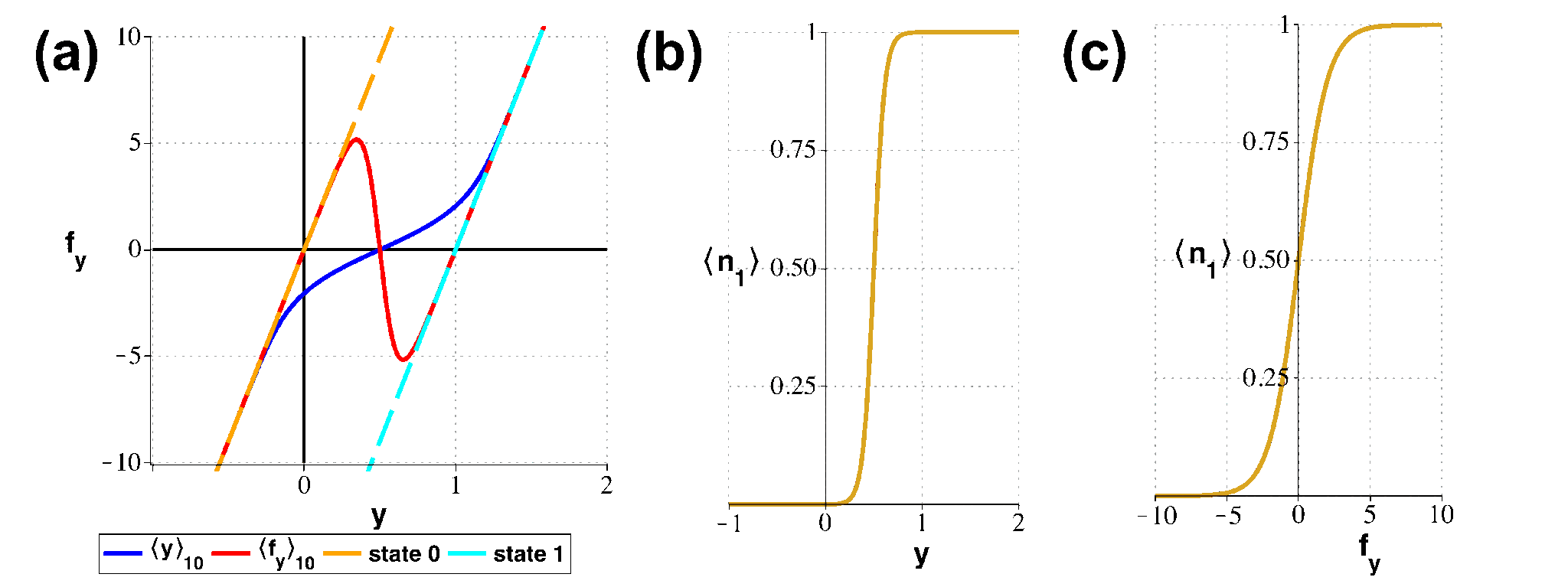}}
\caption{Transverse bending-force response of a grafted two-state semiflexible polymer with reversible spontaneous curvature. (a) The blue and red curves indicate the average displacement $\langle y \rangle_{01}$ and the average force $\langle f_y \rangle_{01}$ for the two-state system, respectively. The dashed orange and cyan lines represent the forces for the uncurved state 0 and the curved state 1, respectively. (b) Probability of the curved state, $\langle n_1 \rangle_H$, plotted against the displacement $y$. (c) Probability of the curved state, $\langle n_1 \rangle_G$, plotted against the bending force $f_y$. The calculation uses the following parameters: $\beta=1$, $\epsilon=0$, $s=L=1$, $c=2$, $L_p^{(0)}=L_p^{(1)}=12$.}
\label{fig:FY_c}
\end{figure}
\begin{figure}[h]
\centering{\includegraphics[width=0.5\textwidth]{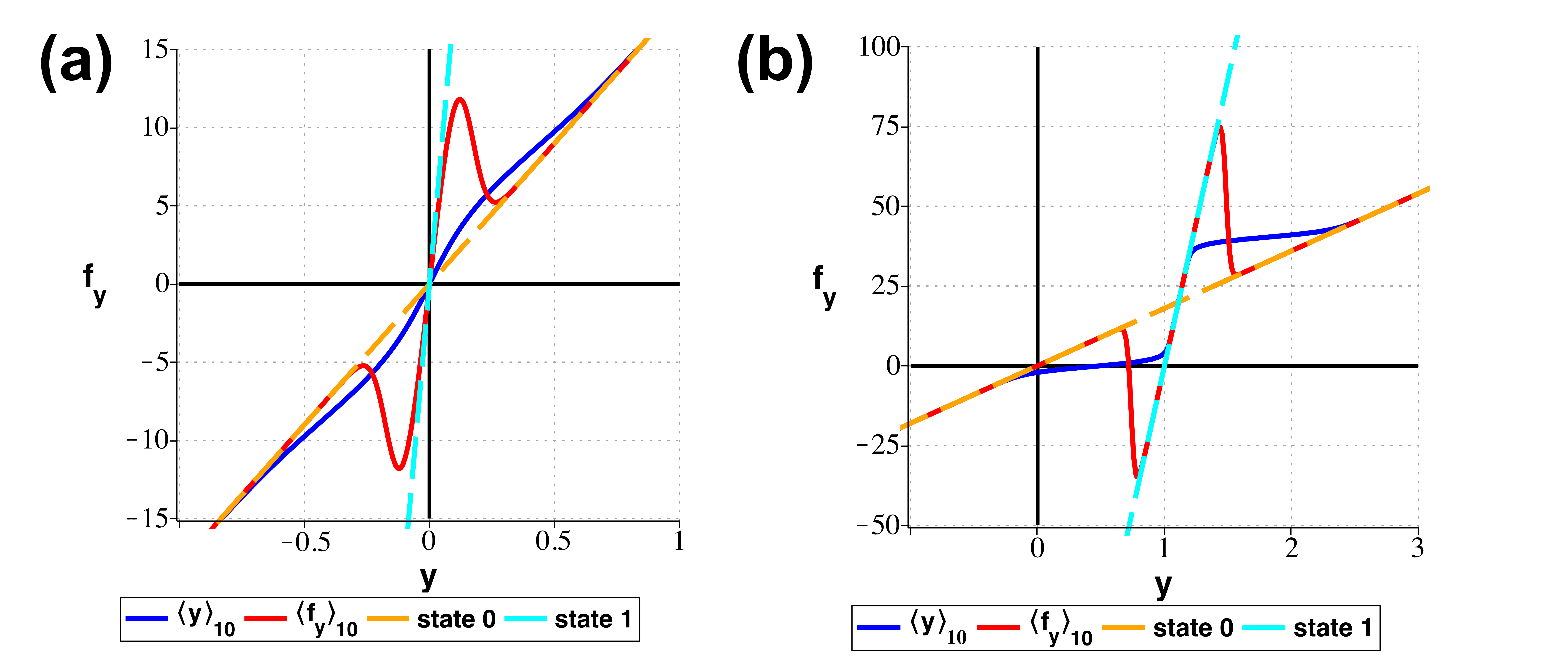}}
\caption{Transverse force–displacement responses for two-state grafted polymers. (a) Persistence length switching only, with \(c=0\). (b) Simultaneous switching of persistence length and spontaneous curvature, with \(c=2\). The blue curves denote the Gibbs response \(\langle y\rangle_{01}\), and the red curves denote the Helmholtz response \(\langle f_y\rangle_{01}\). The orange and cyan dashed lines are the single-state branches for states 0 and 1, respectively. Parameters: $\beta=1$, $\epsilon=0$, $s=L=1$, $L_p^{(0)}=12$, $L_p^{(1)}=120$.
}
\label{fig:FY_b_bc}
\end{figure}

When the controlled displacement is the physical lab-frame transverse tip displacement \(y\), spontaneous curvature shifts the preferred displacement of the curved state. Consequently, even when the two states have the same bending stiffness, the internal-state occupation depends on \(y\) in the Helmholtz ensemble and on \(f_y\) in the Gibbs ensemble, so that the two averaged force--displacement relations are not simple inverses of one another. As shown in Fig.~\ref{fig:FY_c}, the fixed-force response remains monotonic, whereas the fixed-displacement response can exhibit softening or negative differential stiffness near the state crossover, consistent with the general mechanism discussed in Sec.~\ref{sec:SP_const}. The relatively large value of \(cs\) is used only to make the state-switching response visually clear. Quantitative use of the weak-bending propagator should be restricted to \(cs\ll 1\), although the displaced-harmonic mechanism itself is not expected to depend on this particular parameter choice.

When spontaneous curvature and bending stiffness switch simultaneously, their competition can additionally produce the reentrant response shown in Fig.~\ref{fig:FY_b_bc} (b). For the parameters considered, a transverse force applied in the direction of the preferred curvature favors the curved, stiffer state at intermediate loading, whereas stronger bending eventually favors the softer state because of its larger transverse compliance. The resulting sequence is therefore from the uncurved, softer state to the curved, stiffer state and then back to the softer state. Unlike tensile loading, which tends to straighten the polymer, a transverse force drives further bending. The portion of Fig.~\ref{fig:FY_b_bc} (b) with \(y\) approaching or exceeding unity should, however, be regarded only as a formal extrapolation: for \(s=L=1\), \(y>1\) is incompatible with the finite contour length, and the weak-bending approximation requires \(y/L\ll1\).

\section{\label{sec:Conclusions}Discussion-Conclusions}

In this work, we investigated the interplay of curvature and the elastic response of two-state semiflexible polymers with reversible spontaneous curvature. We considered semiflexible polymers that fluctuate in an all-or-none fashion between a state with spontaneous curvature and one without. The tensile elasticity of the former is dominated by the unfolding of the spontaneous curvature, whereas that of the latter is dominated by the ironing out of thermal undulations. Unlike previous reversible-chain models based on fluctuating bending stiffness, the bending rigidity was kept fixed in a significant part of our analysis in order to isolate the role of spontaneous curvature. As in several other two-state or multi-state polymers away from the thermodynamic limit, we obtain significant ensemble inequivalence. 

For the tensile elasticity, we used the strong-stretching force--extension relations of a weakly bending filament in two dimensions with prescribed spontaneous curvature as the basis for a minimal two-state construction. Both constant and sinusoidal spontaneous-curvature profiles were considered. In an approximate description, the curvature contribution (without thermal fluctuations) was assigned to a curved branch, while the thermal fluctuation contribution was assigned to an uncurved branch. This approximation allowed us to derive analytic partition functions in both the Gibbs and Helmholtz ensembles, together with the corresponding force--extension relations and curved-state occupation probabilities.

The stretching results show that increasing the applied force suppresses the contribution of spontaneous curvature and shifts the statistical weight toward the uncurved thermal state. Within the ligand-binding interpretation of the model, this corresponds to a force-induced bias against the ligand-bound curved state. For finite two-state systems, the Gibbs and Helmholtz ensembles yield distinct elastic responses. This is a consequence of the fact that the corresponding free energies  cannot be related by a Legendre transformation. In the Helmholtz ensemble, the state switching can also lead to nonmonotonic response and negative differential stiffness, reflecting the ability of the system at fixed extension to redistribute its weight between the two branches. On the other hand, the elastic response in the Gibbs ensemble is always monotonic, as expected from thermodynamics. 

To make our models closer to the experimental reality, we also considered the simultaneous switching of spontaneous curvature and the value of the bending stiffness (higher in the curved state). Interestingly, this implies to two competing tendencies. At lower stretching forces, the curved state is favored. However, in the strong-stretching regime, the state with the larger persistence length is favored. The result is a reentrant crossover between the two states. 

Single-molecule measurements of IHF-induced DNA bending provide a useful experimental analogue. In Ref.~\onlinecite{le2013mechanosensing}, DNA containing a specific IHF binding site showed two extension levels, which were assigned to bent and unbent conformations. As shown in Fig.~2 of that work, the probability of the bent state decreases as the stretching force is raised. Since the observed switching was interpreted as arising mainly from association and dissociation of IHF at the binding site, the decrease in bending probability suggests that stretching makes the bound, bent state less favorable. This trend is consistent with our two-state model, in which the increasing force reduces the occupation of the curved ligand-bound state and shifts the system toward the uncurved state.

We also analyzed a grafted semiflexible polymer subjected to a bending torque \added{or bending force} at its free end. A coarse-grained two-state Hamiltonian, valid in the rod-like (stiff) limit,  was introduced in terms of the tip orientation angle \added{or the tip transverse displacement}, with one uncurved branch and one branch possessing a spontaneous angle \added{or transverse displacement}. The Gibbs ensemble corresponds to fixed applied torque \added{or force}, \added{in the torque-angle system,} which can be physically motivated by a magnetic bead attached to the polymer tip, whereas the Helmholtz ensemble corresponds to fixed tip angle \added{or displacement}. The two ensembles again give different torque--angle responses when both internal states are thermally accessible. 

Taken together, these results demonstrate that reversible spontaneous curvature provides a distinct and minimal route to ensemble inequivalence in semiflexible-polymer elasticity. This mechanism does not require a change in the bending stiffness between the internal states. The predicted ensemble dependence is therefore a consequence of the competition between mechanical deformation and internal state occupation. Our theoretical analysis suggests that the elastic response of curved or ligand-bound semiflexible polymers may depend sensitively on the experimental control protocol, such as fixed force versus fixed extension or fixed torque versus fixed angle. \added{Modern single-molecule experimental spectroscopy techniques are known to be sensitive to external constraints~\cite{neuman2008single,Yao_inequivalence_IntBiol_2025,JeffChen_inequivalence_PRE2025}. Usually, the Gibbs ensemble is more easily realized with magnetic tweezers, whereas the Helmholtz ensemble is more easily realized with atomic force microscopy (AFM). Similar techniques could be used to observe the predicted ensemble inequivalence and non-trivial elasticity.}

Several extensions remain open. The present theoretical work is based on a simplified two-dimensional weak-bending description. A more complete treatment could include three-dimensional curvature and twist, and the kinetics of binding and unbinding under mechanical loading~\cite{wang2019force}. Our investigation focuses on all-or-none transitions. This may be a reasonable approach for hybridization at the oligonucleotide level, but it fails in the case of longer polymers. An arc-length-dependent distribution of occupation numbers would be more appropriate in that case. 

\begin{acknowledgments}

This work was supported by the NRF grant funded by the MSIT of Korea. (RS-2026-25481789). We thank Artur Ugay for interesting discussions and comments.

\end{acknowledgments}

\section*{Data Availability Statement}

The data that support the findings of this study are available from the corresponding author upon reasonable request.

\appendix

\section{\label{appendix:helmholtz_branch}Helmholtz Free Energy of the Spontaneous Constant Curvature Case}

The Helmholtz free energy of the constant-curvature branch is obtained by a branchwise Legendre transformation,
\begin{align}
F_1(x)=G_1(f_1)+f_1x ,
\end{align}
The Gibbs free energy of this branch is
\begin{align}
G_1(f)=-fL-2La f^{-1/2}+2Lb_1 f^{1/2}.
\end{align}
The corresponding force--extension relation is
\begin{align}
x-L=-La f^{-3/2}-Lb_1 f^{-1/2}.
\end{align}
Substituting this relation into the Legendre transformation gives
\begin{align}
F_1(x)
=f(x-L)-2La f^{-1/2}+2Lb_1f^{1/2} \nonumber\\
=-3La f^{-1/2}+Lb_1f^{1/2}.
\end{align}
Using 
\begin{align}
u(x)= f_1(x)^{-1/2}
\end{align}
from Eq.~\eqref{eq:sp_const_f1}, we obtain the compact form
\begin{align}
F_1(x)=-3La\,u(x)+\frac{Lb_1}{u(x)} .
\end{align}

\section{\label{appendix:sinusoidal_helmholtz}
Helmholtz Free Energy of the Spontaneous Sinusoidal Curvature Case}

For the sinusoidal curved branch, the force--extension relation is
\begin{eqnarray}
\frac{x}{L}
=1-\frac{1}{2}\frac{f_qf_c}{(f_q+f)^2}
-b_2 f^{-1/2}.
\end{eqnarray}
Defining
\begin{eqnarray}
\delta\equiv 1-\frac{x}{L},
\qquad
t\equiv \sqrt f,
\end{eqnarray}
we obtain
\begin{eqnarray}
2\delta-\frac{f_qf_c}{(f_q+t^2)^2}-\frac{2b_2}{t}=0.
\end{eqnarray}
Equivalently, \(t\) is the positive root of
\begin{eqnarray}
2\delta t^5-2b_2t^4+4\delta f_q t^3-4b_2f_qt^2 \nonumber\\
+\left(2\delta f_q^2-f_qf_c\right)t-2b_2f_q^2=0 . 
\end{eqnarray}
Since this fifth-order equation does not have a general closed-form solution,
\begin{eqnarray}
f_2(x)=t^2
\end{eqnarray}
is obtained numerically as the positive root. In practice, we solve
\begin{eqnarray}
g(f)=\delta-\frac{1}{2}\frac{f_qf_c}{(f_q+f)^2}
-b_2 f^{-1/2}=0
\end{eqnarray}
by the bisection method. The root is unique because
\begin{eqnarray}
\frac{dg}{df}=\frac{f_qf_c}{(f_q+f)^3}
+\frac{1}{2}b_2 f^{-3/2} >0
\end{eqnarray}
for \(f>0\).

The Helmholtz free energy of the curved branch follows from \(F_2=G_2+fx\). Using
\begin{eqnarray}
G_2(f)=-fL-\frac{L}{2}\frac{f_qf_c}{f_q+f}
+2Lb_2f^{1/2}
\end{eqnarray}
and
\begin{eqnarray}
x-L=-\frac{L}{2}\frac{f_qf_c}{(f_q+f)^2}
-Lb_2f^{-1/2},
\end{eqnarray}
we find
\begin{eqnarray}
F_2=f(x-L)-\frac{L}{2}\frac{f_qf_c}{f_q+f}+2Lb_2f^{1/2} \nonumber\\
=Lb_2f^{1/2}-\frac{L}{2}\frac{f_qf_c(f_q+2f)}{(f_q+f)^2}. 
\end{eqnarray}
Thus, after setting \(f=f_2(x)\),
\begin{eqnarray}
F_2(x)=Lb_2\sqrt{f_2}
-\frac{L}{2}\frac{f_qf_c(f_q+2f_2)}{(f_q+f_2)^2}.
\end{eqnarray}

\section{\label{appendix:three_state}Three State Description}

\begin{figure*}[t]
\centering
\includegraphics[width=\textwidth]{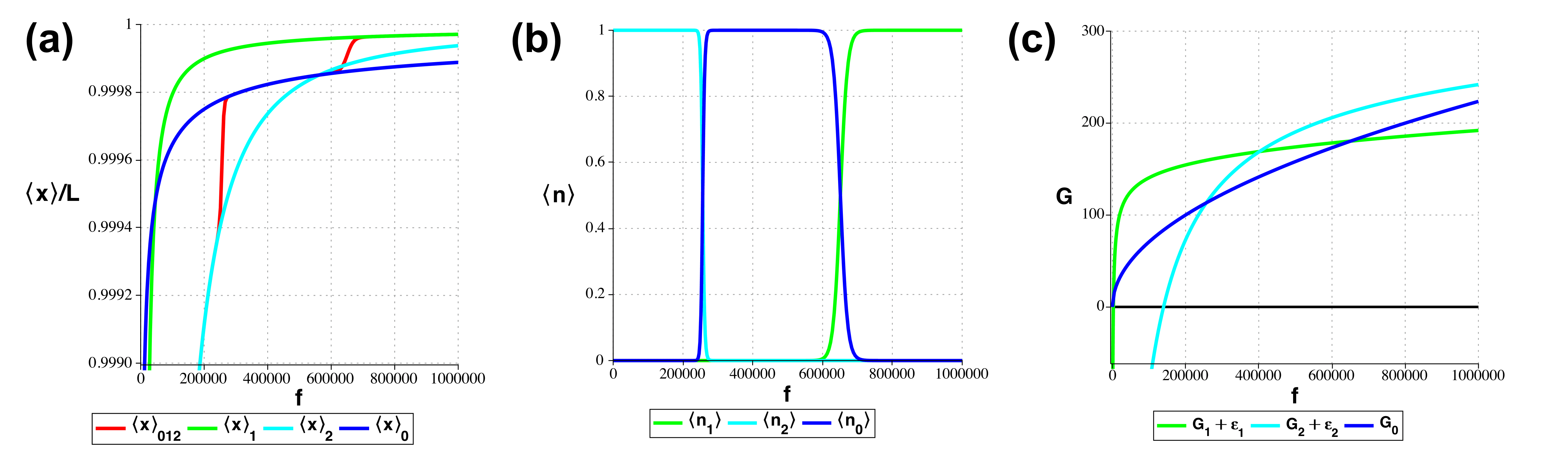}
\caption{Gibbs-ensemble response of a three-state polymer with multiple curvature-generating states. State \(0\) is an uncurved thermal state, state \(1\) is a constant-curvature state, and state \(2\) is a sinusoidal-curvature state. The two curved states are assigned different spontaneous curvatures, \(c_0^{(1)}\) and \(c_0^{(2)}\), to represent distinct curvature-generating mechanisms. (a) Average extension of the three-state system and the three single-state branches. (b) Occupation probabilities of the three states. (c) Effective Gibbs potentials \(\Phi_i^G\) of the three states, including the activation energies of the two curved states. This example represents one possible switching pathway in a multi-state free-energy landscape. Parameters: \(L=1\), \(\beta=1\), \(\epsilon_1=150\), \(\epsilon_2=230\), \(c_0^{(1)}=2\), \(c_0^{(2)}=7\), \(L_p^{(0)}=10\), \(L_p^{(1)}=L_p^{(2)}=200\), and \(q=4\pi\).}
\label{fig:SP_tri_G}
\end{figure*}

In this Appendix, we illustrate how the two-state construction used in the main text can be extended to a multi-state system. We consider three reversible states: an uncurved thermal state \(0\), a constant-curvature state \(1\), and a sinusoidal-curvature state \(2\). The Gibbs partition function is
\begin{eqnarray}
Z_{012}^G(f)= \sum_{i=0}^{2} \exp\left[-\beta \Phi_i^G\right],
\end{eqnarray}
where
\begin{eqnarray}
\epsilon_0 &\equiv 0, \qquad
\Phi_i^G(f) \equiv G_i(f)+\epsilon_i, \quad i\in\{0,1,2\}.
\end{eqnarray}
The ensuing force-extension relation is
\begin{eqnarray}
\langle x\rangle_{012}(f)=\sum_{i=0}^{2}
\langle n_i\rangle_G \langle x\rangle_i ,
\end{eqnarray}
where single-state mean occupation numbers are
\begin{eqnarray}
\langle n_i\rangle_G(f)=\frac{\exp\left[-\beta \Phi_i^G\right]}{\sum_{{\added{j}}=0}^{2} \exp\left[-\beta \Phi_{\added{j}}^G\right]}
\end{eqnarray}
In Fig.~\ref{fig:SP_tri_G} we show the elastic response of this three-state system in the Gibbs ensemble. Although this is a rather academic curiosity, it shows the rich behavior that we can get in the elasticity of those systems. The three-state example in Fig.~\ref{fig:SP_tri_G} represents one possible switching pathway in a multi-state free-energy landscape, rather than a specific biological scenario. In real systems, the selected sequence of states will depend on the curvature geometry, activation energies, and bending stiffnesses of the underlying molecular configurations. Thus, this example illustrates how the present framework can be used to analyze force-dependent state selection among multiple curvature-generating mechanisms.

\section{\label{appendix:bending_force}Grafted Polymer under a Bending \added{Torque or} Force}

Before analyzing the grafted polymer under a bending torque \added{or force}, we consider a grafted semiflexible polymer \added{possessing spontaneous curvature in the rod-like and weak-bending limit.} \deleted{subjected to a transverse bending force at its free end. This case provides a more direct control problem because the conjugate variables are the transverse tip displacement \(y\) and the applied force \(f_y\), rather than the tip angle and torque considered in Sec.~\ref{sec:TA}.}

For a filament with constant spontaneous curvature \(c\), the weak-bending WLC Hamiltonian is
\begin{eqnarray}
H_c[\theta]
=\frac{\kappa}{2}\int_0^s ds'
\left(\frac{d\theta}{ds'}-c\right)^2 .
\end{eqnarray}
Introducing the shifted angle
\begin{eqnarray}
\tilde{\theta}(s)=\theta(s)-\int_0^s c\,ds'=\theta(s)-\added{\theta_c ,\qquad \theta_c\equiv cs},
\end{eqnarray}
the Hamiltonian becomes
\begin{eqnarray}
H_c[\tilde{\theta}]
=\frac{\kappa}{2}\int_0^s ds'
\left(\frac{d\tilde{\theta}}{ds'}\right)^2 .
\end{eqnarray}
\added{Here, \(\theta_c\) is the spontaneous curvature angle at the tip.} Thus, the spontaneous curvature is absorbed into a shift of the angular variable.

Let \(Z(\tilde{\theta},\added{\tilde{x}},\added{\tilde{y}};s)\) denote the configurational partition function for a grafted filament \added{without spontaneous curvature} whose free end has orientation \(\tilde{\theta}\) and position \((\added{\tilde{x}},\added{\tilde{y}})\) at contour length \(s\). The local tangent vector is
\begin{eqnarray}
\mathbf{\added{\tilde{t}}}(s)
=\cos\tilde{\theta}\,\hat{\mathbf e}_{\tilde{x}}
+\sin\tilde{\theta}\,\hat{\mathbf e}_{\tilde{y}} .
\end{eqnarray}
\added{Since the physical angle of a filament with spontaneous curvature is \(\theta(s)=\tilde{\theta}(s)+cs\), in the weak-bending approximation, its configurational degrees of freedom are related to those of a usual filament (tilde symbols) via \(x(s)\simeq\tilde{x}(s)\), \(dy/ds\simeq\tilde{\theta}(s)+cs\), whereas \(d\tilde{y}/ds\simeq\tilde{\theta}(s)\). Hence,}
\begin{eqnarray}
\added{y(s)=\tilde{y}(s)+\int_0^s cs'\,ds'
=\tilde{y}(s)+y_c, \qquad y_c\equiv\frac{cs^2}{2}.}
\end{eqnarray}
\added{Here, \(y_c\) is the transverse tip displacement of the spontaneously curved reference configuration.}

The evolution of \(Z\) along the contour is then described by a  diffusion equation (or imaginary-time \added{Schr\"odinger} \deleted{Schr\"odiger} equation) with a convective
term~\cite{Gompper_Burkhardt,PB_Frey_depinning},
\begin{eqnarray}
\frac{\partial Z}{\partial s}
+\mathbf{\added{\tilde{t}}}\cdot\nabla_{\mathbf{\added{\tilde{r}}}} Z 
=\frac{1}{L_p}\frac{\partial^2 Z}{\partial \tilde{\theta}^2}.
\end{eqnarray}
\added{where \(\tilde{\mathbf r}=(\tilde{x},\tilde{y})\).} In the weak-bending limit, \(|\tilde{\theta}|\ll 1\), we use \(\cos\tilde{\theta}\simeq 1\) and \(\sin\tilde{\theta}\simeq \tilde{\theta}\). The evolution equation reduces to
\begin{eqnarray}
\frac{\partial Z}{\partial s}
+\frac{\partial Z}{\partial \added{\tilde{x}}}
+\tilde{\theta}\frac{\partial Z}{\partial \added{\tilde{y}}}
=\frac{1}{L_p}\frac{\partial^2 Z}{\partial \tilde{\theta}^2}.
\label{eq:bf_evolution}
\end{eqnarray}
To solve Eq.~\eqref{eq:bf_evolution}, we define the Fourier transform of Z as
\begin{eqnarray}
Z(\tilde{\theta},\added{\tilde{x}},\added{\tilde{y}};s)
=\frac{1}{(2\pi)^{3/2}}
\int d^3k\,\hat Z(k_\theta,k_x,k_y;s) \nonumber\\
\times\exp\left[-i(k_\theta\tilde{\theta}+k_x\added{\tilde{x}}+k_y\added{\tilde{y}})\right]. 
\end{eqnarray}
After substituting this form into Eq.~\eqref{eq:bf_evolution} and integrating by parts in \(k_\theta\), the Fourier-space equation becomes
\begin{eqnarray}
\frac{\partial \hat Z}{\partial s}
-ik_x\hat Z
-k_y\frac{\partial \hat Z}{\partial k_\theta}
+\frac{k_\theta^2}{L_p}\hat Z=0 .
\end{eqnarray}
We then use the Gaussian ansatz
\begin{eqnarray}
\hat Z(k_\theta,k_x,k_y;s)
={\cal C}\exp[-a(s)k_\theta^2-b(s)k_\theta k_y \nonumber\\
-g(s)k_y^2-id(s)k_x-ie(s)k_y]
\end{eqnarray}
where $\cal C$ is a constant determined by initial conditions:
\begin{eqnarray}
\lim_{s \to 0} Z = \delta(\tilde{\theta}) \delta(\added{\tilde{x}} - s) \delta(\added{\tilde{y}}).
\end{eqnarray}
Matching the coefficients of equal powers of \(k_\theta\), \(k_x\), and \(k_y\) gives
\begin{eqnarray}
a(s)=\frac{s}{L_p},\qquad
b(s)=\frac{s^2}{L_p},\qquad
g(s)=\frac{s^3}{3L_p}, \nonumber\\
d(s)=-s,\qquad
e(s)=0 .
\end{eqnarray}
Therefore,
\begin{eqnarray}
\hat Z(k_\theta,k_x,k_y;s)
={\cal C} \nonumber\\
\times\exp\left[
-\frac{s}{L_p}k_\theta^2 
-\frac{s^2}{L_p}k_\theta k_y
-\frac{s^3}{3L_p}k_y^2
+ik_xs\right].
\end{eqnarray}
Performing the inverse Fourier transform gives the joint distribution
\begin{eqnarray}
Z(\tilde{\theta},\added{\tilde{x}},\added{\tilde{y}};s)
={\cal C} \,\delta(\added{\tilde{x}}-s) \left(\frac{L_p \sqrt{6\pi}}{s^2}\right) \nonumber\\
\times\exp\left[
-\frac{L_p}{s}\tilde{\theta}^2 
+\frac{3L_p}{s^2}\tilde{\theta}\added{\tilde{y}}
-\frac{3L_p}{s^3}\added{\tilde{y}}^2\right],
\end{eqnarray}
Returning to the original angle \(\theta\), this becomes
\begin{eqnarray}
Z(\theta,x,y;s)
={\cal C}\,\delta(x-s) \left(\frac{L_p \sqrt{6\pi}}{s^2}\right) \nonumber\\
\times\exp\left[-\frac{L_p}{s}(\theta-\added{\theta_c})^2 
+\frac{3L_p}{s^2}(\theta-\added{\theta_c})\added{\left(y-y_c\right)}\right] \nonumber\\
\times\exp\left[-\frac{3L_p}{s^3}\added{\left(y-y_c\right)}^2\right].
\label{eq:bf_joint_distribution}
\end{eqnarray}

\nocite{*}
\bibliography{rSC}

\end{document}